\documentclass[journal=jacsat,manuscript=letter,layout=onecolumn]{achemso}

\usepackage{amsmath}
\usepackage{amssymb}
\usepackage{bm}
\usepackage{booktabs}
\usepackage{graphicx}
\usepackage{url}
\usepackage{microtype}
\usepackage[table]{xcolor}
\usepackage{array}

\newcommand{\Jring}{J_{\mathrm{ring}}}
\newcommand{\Sv}[1]{\mathbf{S}_{#1}}
\newcommand{\sg}[1]{\sigma_{#1}}
\newcommand{\Jt}{\tilde{J}}

\author{Xavier Rocquefelte}
\affiliation{Univ Rennes, CNRS, Institut des Sciences Chimiques de Rennes - UMR 6226, F-35000 Rennes, France}
\email{xavier.rocquefelte@univ-rennes.fr}
\author{Peter Blaha}
\affiliation{Technische Universität Wien, Institute for Materials Chemistry, A-1060 Vienna, Austria}

\title{Mag4: An Automated First-Principles Workflow to Extract
Magnetic Interactions, from Pair Couplings to Four-Spin Ring Exchange}

\abbreviations{DFT,GGA,DMI,SIA}

\begin{document}

\begin{abstract}
Chemists rationalize and design magnetic materials through
structure-property relationships built on pairwise exchange couplings $J$,
which energy-mapping methods now extract routinely from first principles. Yet
whenever four magnetic centers close a loop, as in the CuO$_2$ planes of the
cuprates or the square nets of infinite-layer oxides, a four-spin ring
(cyclic) exchange $\Jring$ can arise, reshaping magnetic ground states and
corrupting the very $J$ values used for design. What has been missing is a
practical way to obtain it: a direct, local extraction of the kind the
four-state method provides for $J$. We supply it by generalizing that method
to a sixteen-state ($2^4$) scheme, and we automate the complete
workflow in the openly available \textsc{Mag4} package, the first automatic
implementation of the four-state family of methods: from a single cif file,
\textsc{Mag4} proposes the supercell that isolates the couplings, generates
the DFT inputs in the exchange-correlation flavour of the user's choice,
extracts $J$ and $\Jring$ with the band-gap and local-moment diagnostics that
certify them, and predicts the magnetic order, its propagation vector, and
the critical temperature. Symmetry reduces the sixteen
configurations to six or eight inequivalent
energies, so the cost is modest. The derivation also shows that the conventional
four-state magnetic coupling, $J$, is itself ring-renormalized, by $\mp 2\Jring S^2$ with the sign
set by the reference state. T-La$_2$CuO$_4$ confirms this quantitatively: three
independent routes agree on $\Jring$ to $0.2\%$, giving $\Jring/J_1 = 0.25$, and a
four-state $J_1$ quoted without naming its reference is wrong by $12\%$ in this
material. The direct sixteen-state extraction itself proves reference-dependent, the
N\'eel and ferromagnetic baths bracketing the mapping value: a fourth-order fingerprint
of interactions beyond the pair-plus-ring model, which additional reference baths
resolve into a bare $\Jring$ and a converging tower of six- and eight-spin loop
couplings. SrFeO$_2$ ($S = 2$), with
the same plaquette yet $\Jring/J = 0.006$, provides the negative control: a plaquette
is necessary for ring exchange, far from sufficient: a
structure-property criterion that chemists can screen against when searching
for new materials hosting strong multi-spin magnetism.
\end{abstract}

\begin{center}
\emph{Dedicated to Myung-Hwan (Mike) Whangbo on the occasion of his 80th birthday.}
\end{center}

\section{Introduction}

Investigating the magnetic properties of solids is like embarking on a journey that
needs more than a lifetime simply to touch, \emph{du bout des doigts}. Along such a
journey, to meet Mike Whangbo and to share his scientific enthusiasm for this field is
one of the precious gifts that life may give.

What has always characterized the Whangbo approach is that the focus falls on the
\emph{driving force}: on the mechanism at the origin of a complex property, rather than
on the number that happens to describe it. As is usual in physics, in chemistry and in
materials science, the interesting problems live at an interface, with complex materials
on one side and sophisticated theories on the other. To answer a scientific question it is
essential to know where the approximations are being made: in the system, in the physics,
and in the method that carries the physics. With Mike, in the spirit of Roald
Hoffmann,\cite{whangbo1978,albright1985} the physics is of prime importance, but the
chemistry stands at exactly the same level. The resulting \emph{jeu d'\'equilibriste}
consists in choosing the best approximants of both, so that the model is just good
enough to meet the outcome it is asked to deliver: to explain, to rationalize, and also
to predict. Numerical accuracy, in this view, is not the point; trends are.

This is Mike's way with magnetism, the distillation of many years spent in the
company of magnetic materials. The spin-dimer
analysis\cite{whangbo2003,dai2001} estimates the exchange couplings of an extended solid
from calculations on the dimer that actually matters, isolating the relevant pair from
the crystal that surrounds it and reading the trend directly from the orbital
interaction. The four-state
method\cite{xiang2011,xiang2013} is the natural descendant of that idea. By flipping the
two spins of a chosen pair through their four relative orientations and combining the
resulting total energies, it \emph{decontaminates} a periodic DFT calculation from the
neighborhood of the selected dimer: every coupling of the pair to the surrounding sea of
spins, and the spin-independent energy itself, cancel identically, and a single $J_{ij}$
survives. The method is exact within the Ising decomposition of the collinear energy and
extends without modification to the full anisotropic exchange tensor.\cite{sabani2020}

The bilinear description that these tools serve is not, however, always
sufficient, and it is the materials themselves that say so.
In the cuprates, polarized-neutron diffuse-scattering measurements on
La$_2$CuO$_4$ gave direct
experimental evidence for four-spin cyclic exchange,\cite{toader2005} and periodic
first-principles calculations subsequently confirmed $\Jring \approx 0.2$--$0.3\,J$ for
several cuprate parents,\cite{moreira2006,calzado2004} where $J$ denotes the
nearest-neighbor Heisenberg exchange, the largest coupling in these materials. The orthorhombic manganites
supply a second and quite different setting: Fedorova \latin{et al.} found that the DFT
energies of many collinear configurations of o-$R$MnO$_3$ simply cannot be fitted by a
Heisenberg-plus-biquadratic model, that adding the four-spin ring term improves the fit
dramatically, and that the resulting coupling, carried by the four-site Mn plaquettes both
within the $ab$ planes and between them, is strong enough to stabilize entirely new
magnetic orders.\cite{fedorova2015,fedorova2018}
For the materials chemist the message of these studies is double: multi-spin
interactions are not exotic corrections but can decide which magnetic order a compound
adopts, and they are attached to a recognizable structural motif, a closed loop of
magnetic centers. The four-spin square studied here is not the only such loop: a
six-spin exchange on the hexagons of the honeycomb lattice has been shown to open the
magnon gap of CoTiO$_3$,\cite{li2024} and in two-dimensional Wigner crystals rings of
two to six electrons compete to select the magnetic ground state, ferromagnetic or
antiferromagnetic.\cite{esterlis2025}
The ring term carries a sting in the
tail, moreover: it is invisible to linear spin-wave theory on the two-sublattice N\'eel
state, so that determining it demands either the paramagnetic structure factor or a
total-energy approach sensitive to the four-spin term.

The physical origin of the ring term is well understood.
Whenever the on-site repulsion $U$ is not overwhelmingly larger than
the hopping $t$, higher-order multi-spin interactions arise in the systematic $t/U$
expansion of the half-filled Hubbard model, derived by Takahashi\cite{takahashi1977} and
given in closed form to fourth order by MacDonald, Girvin and
Yoshioka.\cite{macdonald1988} The leading term of this kind is the four-spin ring (cyclic)
exchange $\Jring$, which describes the coherent circulation of the electrons around a
closed loop of four \emph{coplanar} magnetic centers, each a nearest neighbor of the
next: it appears at order $t^4/U^3$ as the
coefficient of the plaquette operator of Eq.~\eqref{eq:ham} below, accompanied by
corrections of the same order to the bilinear couplings. Whether it matters at all is,
in the first place, a question of local structure. The loop must actually exist: only when the four hops that carry an electron once around
the ring are equivalent do their amplitudes add coherently. Its archetype is a square net
of transition-metal cations bridged by ligands, such as the CuO$_2$ plane of a cuprate,
but any four-site plaquette will serve; where no such loop exists there is no ring
exchange, and even where one does $\Jring$ may be small. Where the loop is present and the
electronic structure cooperates, however, $\Jring$ can be large, contributing both
four-spin and renormalized two-spin terms to the effective Hamiltonian.\cite{toader2005}

What has been missing, then, is not the physics but the tool.
The four-state method does not provide it. The ring operator is quadrilinear in the
spins, so a two-spin flip cannot isolate it: whatever combination of four energies one
forms, the four-spin term either cancels along with everything else or survives mixed
with the pair couplings. What is needed is not a different philosophy but the next member
of the same family.

Here we propose that extension. Flipping the four spins of a nearest-neighbor plaquette
through their $2^4$ collinear arrangements and combining the total energies in a single
signed sum, we obtain a \emph{sixteen-state} method that cancels the spin-independent
constant, the molecular fields exerted by the surrounding bath, and every pair coupling,
leaving $\Jring$ alone. It is the same act of decontamination, performed one order higher.
Symmetry reduces the sixteen configurations to six or eight inequivalent energies,
depending on the stacking of the layers, so the cost remains modest.
Crucially for its adoption, the complete workflow is automated in the openly
available \textsc{Mag4} package:\cite{mag4} from a cif file alone, it proposes the
isolating supercells, generates the DFT inputs in the exchange-correlation flavour of
the user's choice, extracts the couplings with their reliability diagnostics, and
predicts the magnetic order and its critical temperature. Quantities so far reserved
for specialists thus become descriptors that any materials chemist can screen on a
candidate structure, existing or hypothetical.
To test its relevance we compare $\Jring$ extracted by two independent
routes: energy mapping over many collinear configurations, and the sixteen-state method
proposed here. We find, as a by-product of the derivation, that the four-state pair
coupling is itself ring-renormalized, a prediction we confirm quantitatively.

\begin{figure}[htb]
\centering
\includegraphics[width=\linewidth]{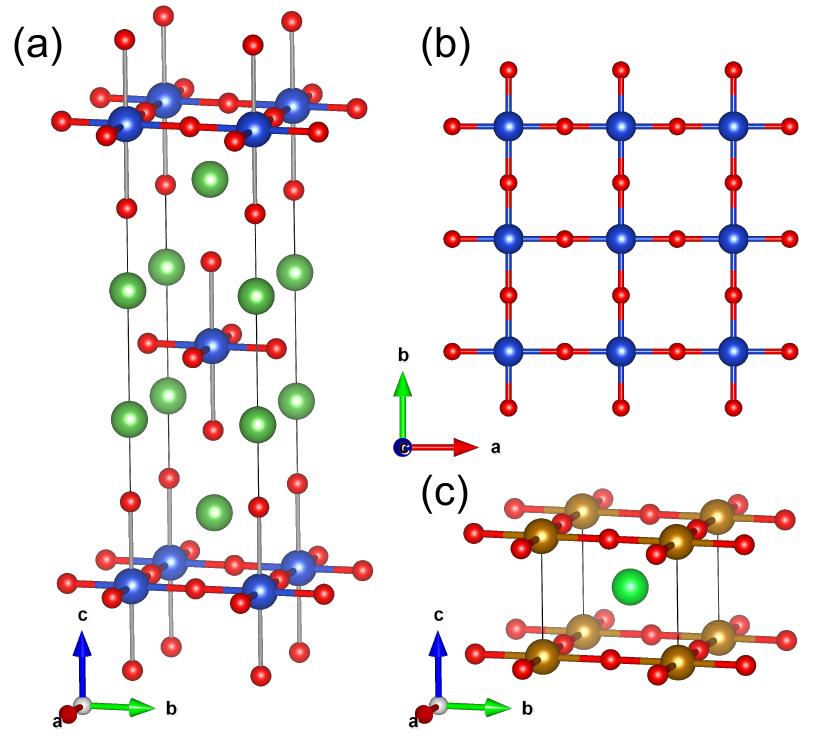}
\caption{The two test materials. (a) T-phase La$_2$CuO$_4$ (K$_2$NiF$_4$ type,
$I4/mmm$; Cu blue, O red, La green). Each Cu sits at the centre of a Jahn--Teller
elongated CuO$_6$ octahedron, the apical oxygen lying along $\mathbf{c}$, and the CuO$_2$
layers are stacked body-centred. (b) A CuO$_2$ layer viewed along $\mathbf{c}$: the Cu
ions form a square lattice bridged by oxygen, so that the nearest-neighbor coupling
proceeds through a $180^{\circ}$ Cu--O--Cu superexchange path. Four such Cu and the four
bridging O define the Cu$_4$O$_4$ plaquette on which the ring exchange of
Fig.~\ref{fig:ring} circulates. (c) Infinite-layer SrFeO$_2$ ($P4/mmm$; Fe gold, O red,
Sr green). Here there is \emph{no} apical oxygen: the Fe(II) is square-planar
coordinated, a coordination unprecedented in an oxide until this compound was
discovered.\cite{tsujimoto2007} The Fe ions nevertheless form the same magnetic square
lattice within each FeO$_2$ plane, but carry $S = 2$ and a very different electronic
structure, which is what makes the pair a demanding test of a single formula.}
\label{fig:structures}
\end{figure}

The method is tested on two materials chosen to be as different as possible, so
that the validation doubles as a structure-property lesson.
La$_2$CuO$_4$ is the archetypal parent of the high-$T_c$ cuprates, a $S = 1/2$
square-lattice antiferromagnet in which ring exchange is known to matter. SrFeO$_2$ is an
infinite-layer iron oxide whose surprising ``flat'' chemistry (square-planar Fe(II) in an
oxide, unprecedented until its discovery\cite{tsujimoto2007}) places $S = 2$ spins on a
square lattice of a completely different electronic character. The two structures are shown in Fig.~\ref{fig:structures}.
Both offer the same magnetic plaquette, and the comparison thereby asks a
structure-property question: is the motif enough? The answer, developed below, is
that it is necessary but far from sufficient, and identifying the chemical factors
that separate the two cases (covalency of the metal-ligand bond, on-site repulsion,
orbital filling) is precisely the kind of criterion needed to discover new materials
hosting strong multi-spin magnetism.


\section{From four states to sixteen states}

\subsection{Collinear reduction of the ring operator}

\begin{figure}[htb]
\centering
\includegraphics[width=\linewidth]{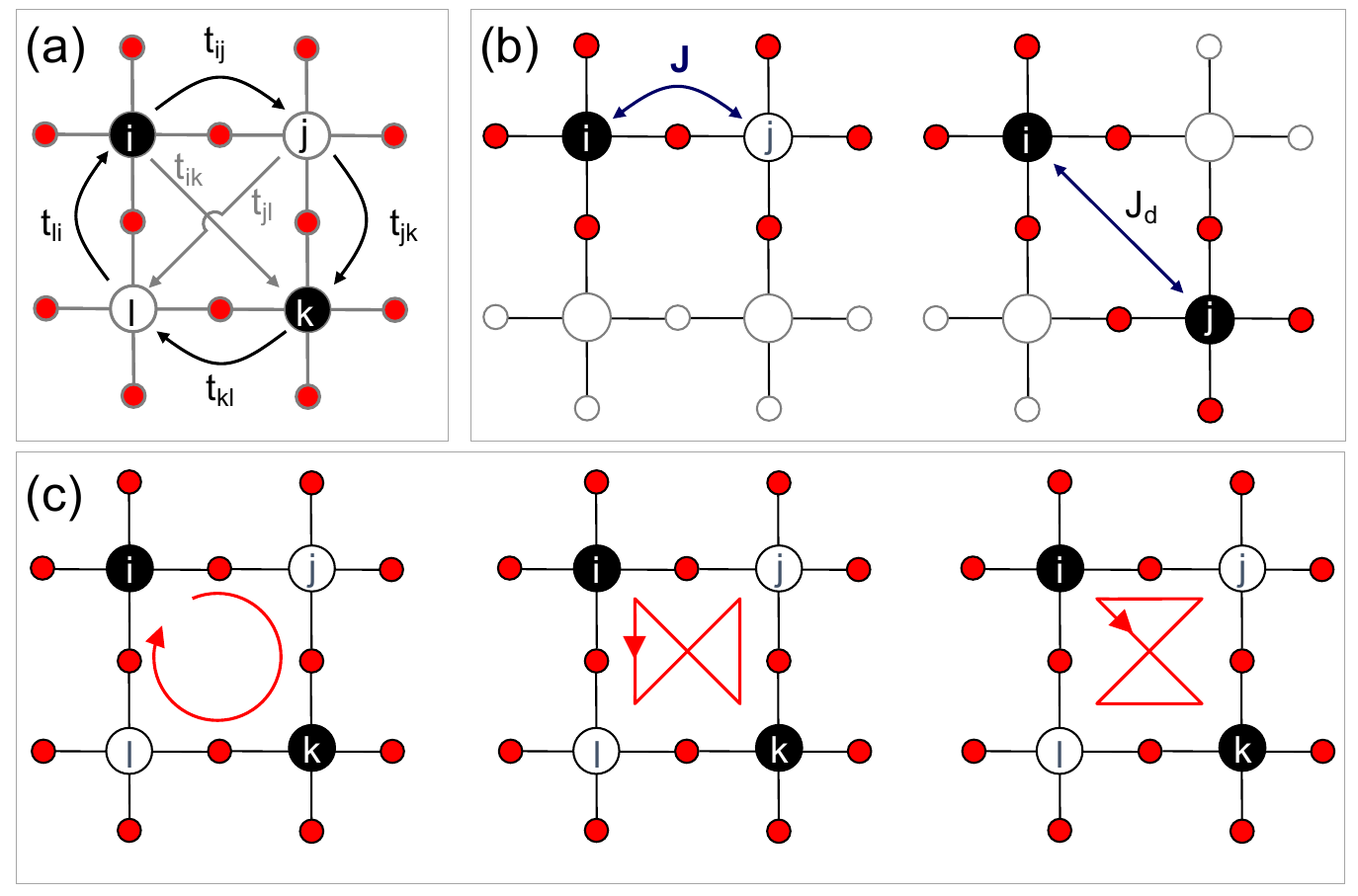}
\caption{Origin and structure of the four-spin ring exchange on a square plaquette.
(a) The four sites $i$, $j$, $k$, $l$ of a nearest-neighbor plaquette, with the
edge hopping integrals $t_{ij}$, $t_{jk}$, $t_{kl}$, $t_{li}$ and the diagonal
hoppings $t_{ik}$, $t_{jl}$ of the underlying Hubbard model; cyclic circulation of the
electrons around this loop generates $\Jring$ at fourth order in the hopping,
$\Jring \propto t^4/U^3$,\cite{takahashi1977,macdonald1988} where $t$ is the transfer integral between neighboring magnetic
sites and $U$ the on-site Coulomb repulsion.
(b) The bilinear couplings retained in Eq.~\eqref{eq:ham}: the nearest-neighbor
exchange $J$ along a plaquette edge and the next-nearest-neighbor exchange $J_d$ along a
plaquette diagonal. (c) The three spin pairings entering the cyclic ring operator of
Eq.~\eqref{eq:ham}: $(\Sv{i}\!\cdot\!\Sv{j})(\Sv{k}\!\cdot\!\Sv{l})$,
$(\Sv{i}\!\cdot\!\Sv{l})(\Sv{j}\!\cdot\!\Sv{k})$ and
$(\Sv{i}\!\cdot\!\Sv{k})(\Sv{j}\!\cdot\!\Sv{l})$, the last entering with a minus sign.
On any collinear configuration all three collapse to the single product
$\sigma_i\sigma_j\sigma_k\sigma_l S^4$ [Eq.~\eqref{eq:lemma}], which is the observation
that makes the sixteen-state extraction possible.}
\label{fig:ring}
\end{figure}

We consider a classical Heisenberg system whose energy is $E = E_0 + E_{\rm
spin}$, with $E_0$ independent of the spin orientations and
\begin{equation}
E_{\rm spin} = \sum_{\langle i,j\rangle} J_{ij}\,(\Sv{i}\!\cdot\!\Sv{j})
\;+\; \sum_{\langle i,j,k,l\rangle} \Jring
\Big[ (\Sv{i}\!\cdot\!\Sv{j})(\Sv{k}\!\cdot\!\Sv{l})
+ (\Sv{i}\!\cdot\!\Sv{l})(\Sv{k}\!\cdot\!\Sv{j})
- (\Sv{i}\!\cdot\!\Sv{k})(\Sv{j}\!\cdot\!\Sv{l}) \Big],
\label{eq:ham}
\end{equation}
where $\langle i,j\rangle$ runs over pairs of spin sites and $\langle i,j,k,l\rangle$
over the four-site nearest-neighbor plaquettes, whose corners are labeled cyclically so
that $(i,j)$, $(j,k)$, $(k,l)$ and $(l,i)$ are the plaquette \emph{edges} and $(i,k)$
and $(j,l)$ its \emph{diagonals} (Fig.~\ref{fig:ring}). We follow the notation of
Fedorova \latin{et al.},\cite{fedorova2018,fedorova2015} whose four-spin coupling
$K_{ijkl}$ is our $\Jring$. In a collinear state with quantization axis $z$ we write
$\Sv{i} = \sg{i} S$ with $\sg{i} = \pm 1$, so that
$\Sv{i}\!\cdot\!\Sv{j} = \sg{i}\sg{j} S^2$. The three products in the bracket of
Eq.~\eqref{eq:ham} then each reduce to $\sg{i}\sg{j}\sg{k}\sg{l} S^4$, two with a
plus sign and one with a minus, so that for \emph{any} plaquette $Q$ of the lattice
\begin{equation}
E_{\rm ring}(Q)\big|_{\rm collinear} = \Jring S^4 \prod_{i \in Q} \sg{i}.
\label{eq:lemma}
\end{equation}
The cyclic operator collapses, on any collinear configuration, to the plain
product of its four spin signs (Fig.~\ref{fig:ring}). This single observation underlies
the whole scheme.

\subsection{Exact decomposition and the extraction formula}

Following the four-state construction, we single out one plaquette with corners
$i,j,k,l$, flip only its four spins, and hold every remaining spin of the supercell
fixed in a reference configuration; these fixed surrounding spins we call the \emph{bath}.
Sorting each term of Eq.~\eqref{eq:ham} by how many of the
four chosen sites it touches gives an exact decomposition,
\begin{equation}
\begin{aligned}
E(\sg{i},\sg{j},\sg{k},\sg{l}) = E_{\rm other}
&+ S\,\big( K_i\sg{i} + K_j\sg{j} + K_k\sg{k} + K_l\sg{l} \big) \\
&+ S^2 \big( \Jt_{ij}\sg{i}\sg{j} + \Jt_{jk}\sg{j}\sg{k}
           + \Jt_{kl}\sg{k}\sg{l} + \Jt_{li}\sg{l}\sg{i} \big) \\
&+ S^2 \big( \Jt_{ik}\sg{i}\sg{k} + \Jt_{jl}\sg{j}\sg{l} \big)
 + \Jring S^4\, \sg{i}\sg{j}\sg{k}\sg{l},
\end{aligned}
\label{eq:decomp}
\end{equation}
in which $E_{\rm other}$ collects everything independent of the four flipped
spins, $K_i$ is the molecular field the bath exerts on site $i$ (the plaquette
counterpart of the $\mathbf{K}_1,\mathbf{K}_2$ of Xiang \latin{et al.}\cite{xiang2011}),
and $\Jt_{ij}$ is an \emph{effective} pair coupling. The decisive property is that all
of these depend on the fixed bath but \emph{not} on $\sg{i},\sg{j},\sg{k},\sg{l}$, so
they are common to all sixteen states. Two features of the coefficients matter for what
follows, and both are derived in full in the Supporting Information. First, the effective
edge couplings are ring-renormalized, $\Jt_{ij}=J_{ij}+\Jring S^2\,\sigma_{e_1}\sigma_{e_2}$,
where $\sigma_{e_1}\sigma_{e_2}$ is the product of the two bath spins of the neighboring
plaquette that shares the edge $(i,j)$, while the two diagonals carry no ring term. Second,
the expansion contains \emph{no} term of odd degree in the flipped spins and only the
chosen plaquette supplies a degree-four term, because two distinct nearest-neighbor
plaquettes share at most an edge; this is the structural reason $\Jring$ can be isolated
cleanly.
Label the sixteen sign patterns $n = 1,\ldots,16$, and let $E_n$ be the energy of pattern
$n$ given by Eq.~\eqref{eq:decomp}. Isolation is achieved by weighting each $E_n$ by its own
ring product $\sg{i}\sg{j}\sg{k}\sg{l}$ and summing. Every lower-degree term cancels
identically ($\sum_n\sg{i}\sg{j}\sg{k}\sg{l}=0$ over the sixteen states, and likewise
against each single spin and each pair), leaving
\begin{equation}
\boxed{\;
\Jring = \frac{1}{16 S^4} \sum_{n=1}^{16}
\sg{i}\sg{j}\sg{k}\sg{l}\, E_n \; }
\label{eq:extract}
\end{equation}
the four-spin counterpart of the four-state formula\cite{xiang2011,sabani2020}
$J_{12} = (E_1 + E_4 - E_2 - E_3)/4S^2$. The prefactor is $16S^4 = 1$ for $S = 1/2$
(La$_2$CuO$_4$), so that $\Jring$ is then simply the signed sum of the sixteen energies,
and $16S^4 = 256$ for $S = 2$ (SrFeO$_2$).

\subsection{Symmetry reduction}

Because the coefficients of Eq.~\eqref{eq:decomp} are common to all sixteen states,
states related by the symmetry of the fixed bath share an energy, and the sixteen
collinear arrangements collapse to few inequivalent \emph{model} energies: six
on the ideal single-layer square lattice, and eight in the body-centered stacking of
T-La$_2$CuO$_4$, where the layers are offset by $(\tfrac12,0,\tfrac12)$ and the
plaquette-centered $C_4$ is not a crystallographic operation. The reduction must be
carried out in the magnetic (grey) group, with time reversal included; the
antiunitary $C_4\times\mathcal{T}$ is what completes the collapse to six on the ideal
lattice. The full class structure, multiplicities, and cancellation identity are given in
the Supporting Information; in practice Eq.~\eqref{eq:extract} is evaluated over the
inequivalent classes alone, each weighted by its ring product and multiplicity, for any
choice of reference bath.

The reduction to six energies is a property of the spin Hamiltonian, not of the
crystal, so the extra configurations left inequivalent by a lower-symmetry cell become a
\emph{test} of the model rather than a cost. In the body-centered supercell, the two extra degeneracies link configurations of
opposite magnetization whose energies the crystal could split only through the interlayer
coupling (negligible here), so our VASP calculations (16 Cu, N\'eel bath, all eight
configurations insulating, gaps 1.10--1.88 eV; Table~\ref{tab:16state}) find the pair
degenerate to 10~$\mu$eV. Details are given in the Supporting Information.

\subsection{Ring exchange contaminates the four-state pair couplings}
\label{sec:contam}

Equation~\eqref{eq:decomp} has a consequence that deserves emphasis. In the
conventional four-state extraction of a nearest-neighbor coupling only the two
spins of the dimer are flipped, so \emph{both} plaquettes containing that bond
have their two remaining corners in the fixed bath. On a N\'eel bath the spins on each of
those corner pairs are antiparallel, so the quantity actually returned is not the bare
coupling but the effective one,
\begin{equation}
J^{\text{4-state}}_{\rm NN}(\text{N\'eel}) = \Jt_{\rm NN} = J_{\rm NN} - 2 \Jring S^2 ,
\qquad
J^{\text{4-state}}_{\rm NN}(\text{FM}) = J_{\rm NN} + 2 \Jring S^2 ,
\label{eq:contam}
\end{equation}
the sign of the renormalization being set by the reference. This is the microscopic
counterpart of the well-known result that cyclic exchange renormalizes the effective pair
exchange entering a spin-wave fit.\cite{toader2005} The shift is $-\Jring/2$ for
$S = 1/2$ and $-8\Jring$ for $S = 2$: far from negligible, so the sixteen-state scheme is
needed not only to obtain $\Jring$ but to correct $J$ itself.

\subsection{A two-bath determination of the ring coupling}
\label{sec:twobath}

The two references in Eq.~\eqref{eq:contam} differ only in the sign of the ring term,
which affords an independent route to $\Jring$ requiring no sixteen-state calculation at
all:
\begin{equation}
\boxed{\;
\Jring = \frac{J^{\text{4-state}}_{\rm NN}(\text{FM})
- J^{\text{4-state}}_{\rm NN}(\text{N\'eel})}{4 S^2} \;}
\label{eq:twobath}
\end{equation}
Two standard four-state calculations with different baths therefore determine the ring
coupling. Because Eq.~\eqref{eq:twobath} involves entirely different energy differences
from Eq.~\eqref{eq:extract}, agreement between the two is a strong internal consistency
check, carried out below. A four-state $J$ reported without its reference bath is therefore ill-defined whenever ring exchange is present.

Equation~\eqref{eq:twobath} carries a second, sharper prediction that could fail. It
applies to the plaquette \emph{edge}. The plaquette \emph{diagonal} behaves differently:
the single plaquette containing both diagonal sites has its two remaining corners as
next-nearest neighbors, whose spins are parallel in a ferromagnetic \emph{and} in a
N\'eel bath alike, so
\begin{equation}
J^{\text{4-state}}_{\rm NNN} = J_{\rm NNN} + \Jring S^2
\qquad\text{for either bath.}
\label{eq:nnn}
\end{equation}
A single pair of four-state calculations therefore tests two things at once: one coupling
must shift by exactly $4\Jring S^2$, and the other must not shift at all.

\subsection{Scope of the method}

Two limitations should be stated explicitly. First, the collinear construction is blind to
the two-site biquadratic term $B_{ij}(\Sv{i}\!\cdot\!\Sv{j})^2$, which reduces to a
spin-independent constant on any collinear state; for the isolation of $\Jring$ this is an
advantage, but a biquadratic coupling, if required, must be obtained separately from
non-collinear states. Second, the scheme returns the \emph{diagonal} matrix element of the
ring operator in the broken-symmetry (Ising) sense of Moreira, Calzado and
Malrieu;\cite{moreira2006,calzado2004} the off-diagonal flip-flop parts of the
cyclic-permutation operator are invisible to collinear DFT. This is precisely what is
required to extract the coefficient $\Jring$ of Eq.~\eqref{eq:ham}, but it is not the same
object as the Dirac permutation amplitude, as discussed with the results.

\section{Computational details}

Spin-polarized total energies were computed with the projector augmented-wave
method\cite{blochl1994} as implemented in VASP,\cite{kresse1996} using the PBE
form of the generalized-gradient approximation\cite{pbe1996} with a Hubbard
correction (GGA$+U$)\cite{dudarev1998} on the transition-metal $d$ states. A
plane-wave cutoff of 500~eV and an energy convergence threshold of $10^{-5}$~eV
were used throughout. Brillouin-zone integrations for the SrFeO$_2$ energy mapping used
the tetrahedron method with Bl\"ochl corrections (\texttt{ISMEAR} $= -5$); the
per-configuration band gaps and local moments quoted as diagnostics below were verified
to be unchanged under Gaussian smearing (\texttt{ISMEAR} $= 0$, $\sigma = 0.01$~eV).

For La$_2$CuO$_4$ we adopt the ideal tetragonal (K$_2$NiF$_4$, $I4/mmm$) T-phase cell,
$a = 3.7817$~\AA, with $U_{\rm eff} = 8$~eV on Cu; the ring plaquette is the Cu$_4$O$_4$
square, whose edges are the nearest-neighbor bonds (3.7817~\AA) and whose diagonals are
the next-nearest-neighbor bonds (5.3481~\AA). For SrFeO$_2$\cite{tsujimoto2007} we use the
$P4/mmm$ infinite-layer cell ($a = 5.6356$~\AA, $c = 3.4580$~\AA), with
$U_{\rm eff} = 4$~eV on Fe; the ring plaquette is the in-plane Fe$_4$ square of edge
3.9850~\AA, whose diagonal is 5.6356~\AA. The magnetic moments correspond to $S = 1/2$
(Cu$^{2+}$) and $S = 2$ (Fe$^{2+}$).

Several supercells were used, and each is named with the result it produced; they were
chosen so that the probed dimer or plaquette is isolated from its periodic images beyond
the longest coupling retained, a requirement discussed further below. Two cell shapes
recur: direct repetitions of the crystallographic cell, and repetitions of the
$\sqrt2 a \times \sqrt2 a \times c$ cell, whose in-plane vectors run along the diagonals
of the magnetic square lattice, at $45^{\circ}$ to the crystallographic axes, and which
holds two magnetic sites per layer. We refer to the latter as the $\sqrt2$ cell; it is
the natural cell of the N\'eel order, and its repetitions isolate a plaquette with fewer
atoms than repetitions of the crystallographic cell.

The decomposition and extraction are written out explicitly for the
SrFeO$_2$ coupling topology in the Supporting Information: there the plaquette lies in
the FeO$_2$ plane, while $J_1$ (along $\mathbf{c}$) and $J_3$ (the inter-layer
diagonal) connect the plaquette only to the bath and cancel identically in the
extraction.

\subsection{Implementation: the MAG4 package}
\label{sec:mag4}

The obstacle to using these methods is rarely the physics, but the chain of practical
steps that surrounds it. One must
identify the magnetic sublattice and its coupling shells, choose a supercell large enough
to isolate the probed dimer or plaquette from its periodic images, enumerate the
symmetry-inequivalent spin configurations together with their multiplicities, write the
corresponding inputs, and then reduce the resulting total energies with the right signs and
weights. Each step is elementary and each is easy to get wrong, as the discussion of
supercell aliasing below illustrates.

All the calculations reported here were prepared and analysed with \textsc{Mag4}, a Python
suite we have developed for this purpose.\cite{mag4} Starting from a crystallographic
information file, it determines the magnetic sublattice and the inequivalent exchange paths
by symmetry, builds the supercell, and generates the spin configurations required for
energy mapping, for the four-state method, and for the sixteen-state method introduced
here, reducing them with the magnetic (grey) group so that time reversal is included. It
writes ready-to-run inputs for either VASP or WIEN2k, and afterwards parses the outputs,
checks that every configuration is converged and carries the intended local moments, and
returns the couplings, either by the closed-form combination of Eq.~\eqref{eq:extract}
(evaluated over the inequivalent classes, each weighted by its ring product and
multiplicity) or by a least-squares fit over many configurations, with residuals and error bars. The figures of merit quoted throughout this paper, the fit root-mean-square (RMS) residual and the per-configuration residuals of Fig.~\ref{fig:sfo} among them, are produced by it.
\textsc{Mag4} is openly available.\cite{mag4} A full description of the package, of its
symmetry analysis and of the additional interactions it can treat will be published
separately; here we record only that the results below were obtained with it, and that the
sixteen-state method requires no more effort from the user than the four-state method it
extends.

\paragraph{Gap and moment as validity diagnostics.}
A final, practical point belongs with the package, because it governs when the couplings it returns can be trusted. Every one of these methods, energy mapping, the four-state method and the sixteen-state method, rests on the assumption that each configuration entering the extraction remains in the localized-spin regime that Eq.~\eqref{eq:ham}
describes. The two quantities that decide whether it does, the band gap and the local moment, are obtained for free in any total-energy calculation, and \textsc{Mag4} prints both for every configuration, flagging any that is metallic or whose magnetization is inconsistent with its intended pattern. Of the two the gap is the more incisive: a configuration can keep a healthy local moment yet cross into the gapless regime, where a Heisenberg mapping no longer applies, so a moment check alone can give false reassurance.
The gap and the moment should therefore accompany any extracted coupling as a validity certificate, not be left implicit.

This carries a direct consequence for the choice of reference. Because the gap falls with the ferromagnetic character of a configuration, a ferromagnetic reference operates the calculation closest to the metallic edge and is the first to cross it as correlations weaken. For gap-sensitive materials the couplings $J$ and $\Jring$ are therefore best extracted with the four-state and sixteen-state methods anchored on the ground-state-like reference, which keeps the largest gap and the best-conserved moments, rather than on a ferromagnetic reference or on an energy mapping forced to sample high-energy
configurations of uncertain character. The local methods are in this sense the safer
instrument: they never have to leave the neighbourhood of the ground state, where the localized-spin model is secure. La$_2$CuO$_4$ and SrFeO$_2$ illustrate the two regimes; the supporting gaps, moments and couplings are given with the results below.

\section{Results and discussion}

\begin{figure}[htb]
\centering
\includegraphics[width=\linewidth]{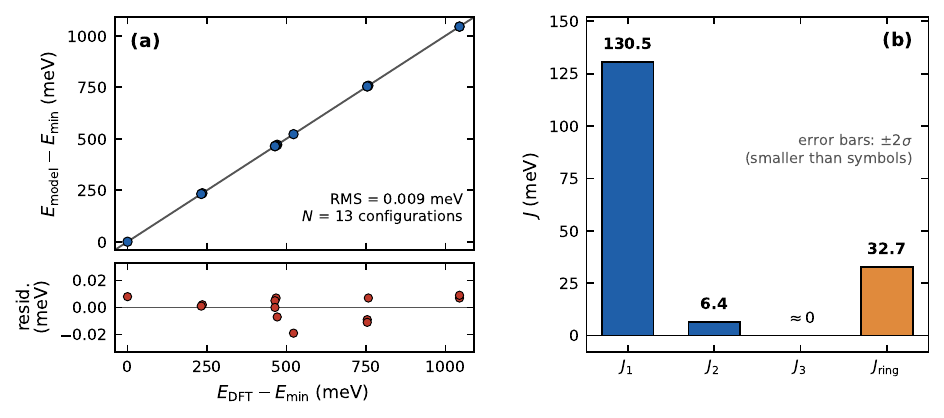}
\caption{Energy mapping for T-La$_2$CuO$_4$ in the $2\times2\times1$ supercell.
(a) Model energies against DFT energies for the 13 inequivalent collinear
configurations, both measured from the lowest-energy state. The line is $y = x$, not a
fit to the points. The residuals, plotted below on a scale four orders of magnitude
finer, lie within $\pm 0.02$~meV and give an RMS of $0.009$~meV over an energy range of
more than $1$~eV, so the spin model of Eq.~\eqref{eq:ham} reproduces the first-principles
energies essentially exactly.
(b) The couplings extracted from that fit: the nearest-neighbor exchange $J_1$, the
next-nearest-neighbor $J_2$, the interlayer $J_3$, and the four-spin ring coupling
$\Jring$. Error bars are $\pm 2\sigma$ from the least-squares fit and are smaller than
the symbols. Two features are worth noting: $J_3$ is indistinguishable from zero,
confirming that the interlayer coupling is negligible; and $\Jring$ is not a small
correction but the second-largest term in the Hamiltonian, five times $J_2$ and one
quarter of $J_1$.}
\label{fig:mapping}
\end{figure}

\paragraph{La$_2$CuO$_4$: bilinear couplings and ring exchange.}
Couplings were obtained from two independent supercells (Table~\ref{tab:lco}). In a
$2\times2\times1$ cell (8 Cu, 13 configurations, $9\times9\times5$ $k$-mesh) the
energy-mapping fit has RMS $= 8.6$~$\mu$eV and gives $J_1 = 130.51$~meV,
$J_2 = 6.41$~meV, a negligible interlayer term, and $\Jring = 32.69$~meV. The quality of
that fit, and the resulting couplings, are shown in Fig.~\ref{fig:mapping}. In a
$2\times3\times1$ cell (12 Cu, 49 configurations) the same analysis gives
$J_1 = 129.89$, $J_2 = 6.66$ and $\Jring = 32.32$~meV. Two cells of different shape,
with different $k$-meshes, therefore agree on the dimensionless ratio to better than
$1\%$:
\begin{equation}
\Jring/J_1 = 0.2505 \;\;(2\times2\times1), \qquad 0.2488 \;\;(2\times3\times1),
\label{eq:ratio}
\end{equation}
in close agreement with the periodic-DFT result of Moreira \latin{et
al.},\cite{moreira2006} whose values are listed alongside ours in Table~\ref{tab:lco}.
Their hybrid-functional calculation gives $J = 140.1$ and $\Jring = 35.8$~meV, both
$7$--$9\%$ above our GGA$+U$ values, a shift of the expected size and direction between
the two exchange-correlation treatments; the small diagonal coupling, more delicate as
discussed with the shell aliasing below, differs more ($J_2 = 8.8$ against our
$6.4$~meV). The dimensionless ratio, however, agrees to $2\%$: $0.250$ here against
their $0.256$. Meanwhile $\Jring = 32.7$~meV lies within the
experimental estimates of Coldea \latin{et al.}\cite{coldea2001} ($38 \pm 8$~meV) and
Mizuno \latin{et al.}\cite{mizuno1998} (40~meV). The experimental ratio itself
spreads with temperature and with the model used in the fit: the Hubbard-expansion
analysis of Coldea \latin{et al.}\cite{coldea2001} gives $\Jring/J_1 = 0.27$ at
295~K but $0.42$ at 10~K, and the analysis of Toader \latin{et
al.}\cite{toader2005} reaches $\approx 0.5$. As Moreira \latin{et
al.}\cite{moreira2006} observed of their own, nearly identical ratio, values near
$0.25$ are in close agreement with the generally accepted
$\Jring/J \approx 0.3$ for this compound, and our two supercells sit in the same
place.

\begin{table}[htb]
\caption{Exchange couplings of T-phase La$_2$CuO$_4$ (meV) from two supercells. The Moreira \latin{et al.}\cite{moreira2006} column is their periodic hybrid-functional (Fock-35, \textsc{Crystal}) result on the experimental structure, with the same ring operator and broken-symmetry mapping; their $J$, $J_d$, $\Jring$ are our $J_1$, $J_2$, $\Jring$. The experimental column is the $t$--$U$ Hubbard spin-wave fit of Coldea \latin{et al.}\cite{coldea2001} at 295~K, their $J' = J''$ mapping to our $J_2$ and $J_4$; the same fit at 10~K gives $J = 146.3 \pm 4$ and $\Jring = 61 \pm 8$~meV ($\Jring/J_1 = 0.42$).}

\label{tab:lco}
\centering
\small
\begin{tabular}{l c r r r r}
\toprule
Coupling & $d$ (\AA) & $2\times2\times1$ & $2\times3\times1$ & Ref.~\citenum{moreira2006}
& Exp.\cite{coldea2001} \\
\midrule
$J_1$ (NN, plaquette edge)      & 3.7817 & 130.51  & 129.89 & 140.1 & $138.3 \pm 4$ \\
$J_2$ (NNN, plaquette diagonal) & 5.3481 & 6.41    & 6.66   & 8.8   & $2.0 \pm 0.5$ \\
$J_3$ (interlayer)              & 7.1441 & $-0.001$& $-0.014$ & & \\
$J_4$ (3rd in-plane)            & 7.5634 & indet.\ & $2.90$ & & $2.0 \pm 0.5$ \\
$\Jring$                        & ---    & 32.69   & 32.32  & 35.8 & $38 \pm 8$ \\
\midrule
$\Jring/J_1$                    &        & 0.2505  & 0.2488 & 0.256 & 0.27 \\
fit RMS (meV)                   &        & 0.0086  & 0.697  & & \\
\bottomrule
\end{tabular}
\end{table}

\paragraph{Quantitative confirmation of the ring renormalization of $J$.}
Equation~\eqref{eq:contam} makes a sharp, falsifiable prediction: a four-state
calculation performed on a N\'eel reference must return not the true $J_1$ but
$J_1 - 2\Jring S^2$. We test it directly. In the \emph{same} $2\times2\times1$ cell and
at the \emph{same} $9\times9\times5$ $k$-mesh as the reference mapping run, the
conventional four-state method returns $J_1^{\text{4-state}}(\text{N\'eel}) =
114.19$~meV, a deficit of $16.31$~meV below the mapping value $J_1 = 130.51$~meV.
Inverting Eq.~\eqref{eq:contam}, that deficit is an independent determination of the
ring coupling that uses only bilinear energy differences,
\begin{equation}
\Jring = \frac{J_1 - J_1^{\text{4-state}}(\text{N\'eel})}{2S^2}
       = \frac{16.31}{0.5} = 32.62~\text{meV},
\label{eq:jring_from_contam}
\end{equation}
in agreement with the $32.69$~meV of the energy mapping to $0.2\%$. This is a stringent,
parameter-free check of the entire framework, obtained without any sixteen-state
calculation: the four-state coupling is ring-renormalized by exactly the predicted
amount, and a four-state $J$ quoted without reference to its bath is, in a material with
ring exchange, wrong by $12\%$ here.

\paragraph{Two independent determinations of $\Jring$ from bilinear energies.}
Equation~\eqref{eq:twobath} can now be applied directly. Repeating the four-state
calculation of the nearest-neighbor bond in the same supercell and $k$-mesh, but on a ferromagnetic reference, returns
$J_1^{\text{4-state}}(\text{FM}) = 146.83$~meV, against
$J_1^{\text{4-state}}(\text{N\'eel}) = 114.19$~meV. Since $4S^2 = 1$ for $S = 1/2$, the
difference \emph{is} the ring coupling:
\begin{equation}
\Jring = J_1^{\text{4-state}}(\text{FM}) - J_1^{\text{4-state}}(\text{N\'eel})
= 146.83 - 114.19 = 32.64~\text{meV}.
\label{eq:twobath_result}
\end{equation}
Three routes now give the same number from three different sets of energy differences:
$32.69$~meV from the energy mapping, $32.62$~meV from inverting the deficit against the
mapping $J_1$ [Eq.~\eqref{eq:jring_from_contam}], and $32.64$~meV from the two baths
[Eq.~\eqref{eq:twobath_result}]. The spread is $0.07$~meV, or $0.2\%$ of $\Jring$. Applying
the correction from either side likewise recovers the true coupling, and brackets it:
$146.83 - 2\Jring S^2 = 130.49$~meV and $114.19 + 2\Jring S^2 = 130.54$~meV, against
$130.51$~meV from the energy mapping. The Supporting Information collects, as a
practical recipe, the ring correction to apply to the four-state couplings for every
supercell and reference bath used here.

The same pair of calculations tests the second, more delicate prediction,
Eq.~\eqref{eq:nnn}. The next-nearest-neighbor coupling should be \emph{unaffected} by the
choice of reference, because the spins on the two bath corners flanking the probed
diagonal (themselves next-nearest neighbors, on the same sublattice) are parallel in a
N\'eel bath as in a ferromagnetic one. The two calculations return $J_2^{\text{4-state}}(\text{FM}) = 14.582$~meV and
$J_2^{\text{4-state}}(\text{N\'eel}) = 14.588$~meV: a difference of $0.006$~meV, where the
nearest-neighbor coupling in the very same calculations shifts by $32.64$~meV. One coupling
moves by exactly $\Jring$ and the other does not move at all, precisely as
Eqs.~\eqref{eq:twobath} and \eqref{eq:nnn} require.

The absolute value should not be
misread, however: $14.58$~meV is the \emph{effective} diagonal coupling, containing the
reference-independent ring shift $+\Jring S^2 = 8.17$~meV of Eq.~\eqref{eq:nnn}.
Subtracting it recovers the bare $J_2 = 6.41$ and $6.42$~meV on the two references, both
matching the energy-mapping value; taken at face value an uncorrected four-state result
would overestimate the next-nearest-neighbor exchange by more than a factor of two. This
is the sharpest confirmation of the framework that we have, and it costs eight
self-consistent calculations.

\begin{table}[htb]
\caption{$\Jring$ for T-La$_2$CuO$_4$ from three independent routes, all in the same
$2\times2\times1$ cell at the same $9\times9\times5$ $k$-mesh, and the reference-dependence
test of the pair couplings. $S = 1/2$, so $4S^2 = 1$ and $2S^2 = 1/2$.}
\label{tab:routes}
\centering
\small
\begin{tabular}{l l r}
\toprule
Route & Energy differences used & $\Jring$ (meV) \\
\midrule
energy mapping & 13 collinear configurations, global fit & 32.69 \\
one-bath, Eq.~\eqref{eq:jring_from_contam}
  & $\big[J_1 - J_1^{\text{4-state}}(\text{N\'eel})\big]/2S^2$ & 32.62 \\
two-bath, Eq.~\eqref{eq:twobath}
  & $\big[J_1^{\text{4-state}}(\text{FM}) - J_1^{\text{4-state}}(\text{N\'eel})\big]/4S^2$
  & 32.64 \\
\midrule
\multicolumn{2}{l}{spread} & 0.07 \\
\bottomrule
\end{tabular}

\vspace{6pt}
\begin{tabular}{l r r r}
\toprule
Four-state coupling & N\'eel ref. & FM ref. & difference \\
\midrule
$J_1^{\text{4-state}}$ (plaquette edge), must shift by $4\Jring S^2 = \Jring$
  & 114.19 & 146.83 & $+32.64$ \\
$J_2^{\text{4-state}}$ (plaquette diagonal), must not shift [Eq.~\eqref{eq:nnn}]
  & 14.588 & 14.582 & $-0.006$ \\
\bottomrule
\end{tabular}
\end{table}

\paragraph{Sixteen-state extraction: a genuine reference dependence.}
The sixteen-state extraction was performed on a body-centered supercell of sixteen Cu, eight per CuO$_2$ layer (a $2\times2\times1$ repetition of the $\sqrt2$ cell, 112 atoms, $6\times6\times5$ $k$-mesh) in which the probed plaquette is fully isolated: no periodic image shares an
edge or a corner with it, so the fixed-bath decomposition of
Eq.~\eqref{eq:decomp} applies without image contamination. Every
diagnostic of the preceding sections is clean. The grey group reduces the sixteen
states to eight configurations on the N\'eel reference and six on the ferromagnetic
one; all are insulating, with gaps of $1.10$ to $1.88$~eV (N\'eel) and $0.58$ to
$0.60$~eV (FM); the local moments are healthy on every site, $|\mu| = 0.74 \pm 0.03\,\mu_B$ on the N\'eel
reference and $0.80 \pm 0.03\,\mu_B$ on the ferromagnetic one; and the two degeneracies
that the spin model predicts beyond the grey group are satisfied to $10$~$\mu$eV. The
two references nevertheless return different couplings (Table~\ref{tab:16state}):
$\Jring = 27.54$~meV on the N\'eel bath against $34.52$~meV on the ferromagnetic one,
each value identical to that obtained in the smaller eight-Cu cell. The
all-electron WIEN2k code\cite{blaha2020}, run on the same supercell at matched $k$-mesh density,
reproduces the dependence in full, $23.45$~meV on the N\'eel bath against
$36.35$~meV on the ferromagnetic one, with every configuration again insulating
(gaps of $0.48$--$0.50$~eV on the ferromagnetic reference and
$1.02$--$1.85$~eV on the N\'eel one, close to their VASP counterparts): the
effect is not an artifact of the pseudopotential construction. Each number
is a small residual of near-cancelling energies, $0.4\%$ of the configuration-energy
spread, so this double supercell-robustness is itself significant: the difference is
not noise.

Nor is it an artifact of the numerics: replacing the tetrahedron scheme by Gaussian
smearing and tightening the electronic convergence to $10^{-8}$~eV moves both values by a
few $\mu$eV only. By construction Eq.~\eqref{eq:extract} is reference-independent for the
Hamiltonian of Eq.~\eqref{eq:ham}, so a dependence that survives the isolation of the
plaquette, the gap and moment diagnostics, and the grey-group analysis is a genuine
signal that the spin Hamiltonian of La$_2$CuO$_4$ contains interactions beyond the
pair-plus-ring model. Its character is clear from where it appears. The bilinear sector
is clean: $J_2$ shifts by $0.006$~meV between the two references, and the three routes
resting on bilinear energy differences agree with the mapping to $0.2\%$. The
contamination is confined to the quadrilinear channel, as expected from the six-spin
loops of order $t^6/U^5$ that extend over the plaquette and its bath: in the
$\chi$-weighted sum of Eq.~\eqref{eq:extract} such terms do not cancel but enter
multiplied by products of bath spins, and so change with the reference.

The practical consequence is read off directly: the two references bracket the
energy-mapping value, $27.5 < 32.7 < 34.5$~meV, and the half-splitting of the
sixteen-state pair, $\pm 3.5$~meV (i.e.\ $\pm 11\%$), is the intrinsic accuracy with
which a strictly local quadrilinear probe can define $\Jring$ in a material this strongly
coupled. The same splitting is at the same time a cheap physical diagnostic: obtained from only
two short four-state calculations, the FM--N\'eel difference measures, at fourth order,
how far the material departs from the pair-plus-ring model of Eq.~\eqref{eq:ham}.

\begin{table}[htb]
\caption{$\Jring$ (meV) for T-La$_2$CuO$_4$ from the sixteen-state method,
Eq.~\eqref{eq:extract}, on the isolated-plaquette supercell (16 Cu, $2\times2\times1$
repetition of the $\sqrt2$ cell, $6\times6\times5$ $k$-mesh). $N_{\rm cfg}$ is the grey-group count of inequivalent configurations. Both
VASP values are identical to those obtained in the eight-Cu cell. The WIEN2k
rows are the same supercell (indexed $2\times1\times2$ in the WIEN2k axis
convention) at matched $k$-mesh density: the all-electron code reproduces the
dependence.} 
\label{tab:16state}
\centering
\small
\begin{tabular}{l l c c r}
\toprule
Code & Reference & $N_{\rm cfg}$ & gap range (eV) & $\Jring$ \\
\midrule
VASP   & N\'eel         & 8 & 1.10--1.88 & 27.54 \\
VASP   & Ferromagnetic  & 6 & 0.58--0.60 & 34.52 \\
WIEN2k & N\'eel         & 8 & 1.02--1.85 & 23.45 \\
WIEN2k & Ferromagnetic  & 6 & 0.48--0.50 & 36.35 \\
\midrule
\multicolumn{4}{l}{energy mapping (Table~\ref{tab:lco})} & 32.69 \\
\multicolumn{4}{l}{two-bath, Eq.~\eqref{eq:twobath}}     & 32.64 \\
\bottomrule
\end{tabular}
\end{table}

\begin{figure}[htb]
\centering
\includegraphics[width=\linewidth]{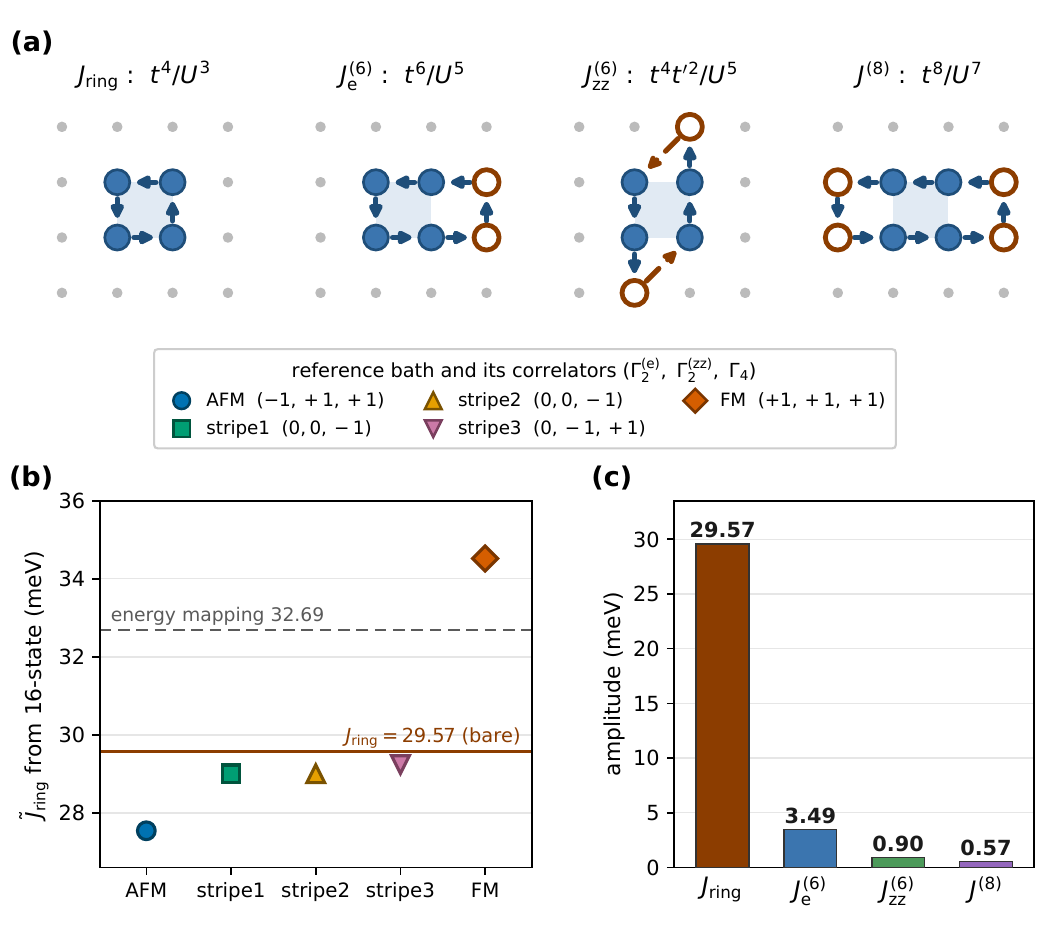}
\caption{Bath decomposition of the sixteen-state extraction for T-La$_2$CuO$_4$.
(a) The loop families of Eq.~\eqref{eq:tower} on the square lattice: plaquette spins
filled, bath spins open; solid bonds are nearest-neighbor hops $t$, dashed bonds
diagonal hops $t'$; the shaded square is the probed plaquette. Of the eight-spin
family only the largest, pure-$t$ loop is shown. (b) The measured
$\Jt_{\mathrm{ring}}(\mathrm{ref})$ for the five reference baths (the legend gives
each bath's correlator triple; AFM denotes the N\'eel bath), against the bare
$\Jring = 29.57$~meV of
Eq.~\eqref{eq:tower} (solid line) and the energy-mapping value (dashed). stripe2
repeats stripe1's flanking signature with the out-of-plane bath Cu reversed, so the
model predicts the two values to coincide; the measurement confirms it to
$1.3$~$\mu$eV. (c) The
solved amplitudes: the bare ring coupling and the three loop-family couplings, a
tower converging by roughly $(t/U)^2$ per order.}
\label{fig:tower}
\end{figure}

\paragraph{Bath decomposition: the reference dependence resolved into couplings.}
If longer loops are the cause, they can be measured, because each enters the
sixteen-state sum multiplied by a \emph{known} function of the bath. The $t/U$ expansion
that yields $\Jring$ at order $t^{4}/U^{3}$ generates on every longer closed loop an
analogous cyclic term of order $t^{\ell}/U^{\ell-1}$, so the leading candidates beyond
the plaquette have $\ell = 6$ and $8$.\cite{takahashi1977,macdonald1988} Adding these
terms to Eq.~\eqref{eq:ham},
\begin{equation}
E_{\rm spin} \;\longrightarrow\; E_{\rm spin}
\;+\; \sum_{\ell = 6, 8, \ldots}\;\sum_{L \in \mathcal{L}_\ell} J_{L}\, O_{L}\,,
\qquad
O_{L}\big|_{\rm collinear} = S^{\ell} \prod_{i \in L} \sg{i}\,,
\label{eq:hfull}
\end{equation}
with $\mathcal{L}_\ell$ the loops of $\ell$ sites and each $O_L$ reducing on collinear
states to the product of its spin signs [generalizing Eq.~\eqref{eq:lemma}]. A loop that
misses a plaquette corner carries at most three plaquette spins and is annihilated by the
$\chi$-weighted sum, as the pair terms are; one that \emph{contains} the plaquette
survives, its four plaquette spins absorbed and its remaining $\ell-4$ bath spins left as
a fixed number of the reference. The method therefore returns exactly
\begin{equation}
\Jt_{\mathrm{ring}}(\mathrm{ref}) \;=\; \Jring
\;+\; S^{2}\!\!\sum_{L_6\,\ni\, i,j,k,l}\!\! J_{L_6}\,
      \big\langle \sg{b_1}\sg{b_2} \big\rangle_{\mathrm{ref}}
\;+\; S^{4}\!\!\sum_{L_8\,\ni\, i,j,k,l}\!\! J_{L_8}\,
      \big\langle \sg{b_1}\sg{b_2}\sg{b_3}\sg{b_4} \big\rangle_{\mathrm{ref}}\,,
\label{eq:jeff}
\end{equation}
an \emph{effective} coupling, tilded like $\Jt_{ij}$ because it is what the probe
measures, not the model parameter. On the square lattice the surviving loops fall into
three families [Fig.~\ref{fig:tower}(a)]: the six-spin \emph{domino} (pure $t$, two bath
spins flanking one edge); the $t'$-assisted six-spin loop whose path zigzags across the
plaquette (two bath spins facing across it on the same sublattice); and the eight-spin
loops, led by the pure-$t$ $3\times1$ strip (four bath spins on two opposite edges), of
which only this largest member is drawn, its $t'$-assisted partners feeding the same
correlator suppressed by $(t'/t)^2$. Each family couples to one bath correlator
$\Gamma_{\ell-4}$: $\Gamma_{2}^{(\mathrm{e})}$ the mean over the four edge-flanking bath
pairs, $\Gamma_{2}^{(\mathrm{zz})}$ over the two across-plaquette pairs, and $\Gamma_{4}$
the product of all four in-plane bath spins. Grouping Eq.~\eqref{eq:jeff} by them gives
the working form
\begin{equation}
\Jt_{\mathrm{ring}}(\mathrm{ref}) \;=\; \Jring
\;+\; \Gamma_{2}^{(\mathrm{e})}\, J^{(6)}_{\mathrm{e}}
\;+\; \Gamma_{2}^{(\mathrm{zz})}\, J^{(6)}_{\mathrm{zz}}
\;+\; \Gamma_{4}\, J^{(8)}\,,
\label{eq:tower}
\end{equation}
each $J^{(\ell)}$ being the spin-scaled sum over the symmetry-equivalent loops of its
family, so that no fragile loop counting enters.

The four unknowns are fixed by the baths the supercell provides. The FM and N\'eel
references have correlator triples $(+1,+1,+1)$ and $(-1,+1,+1)$; two \emph{stripe} baths
zero the correlators in turn, stripe1 with $(0,0,-1)$ and stripe3 with $(0,-1,+1)$. A
fifth, stripe2, repeats stripe1's flanking signature but reverses all eight out-of-plane
bath Cu: no grey-group operation connects them, yet no loop family reaches the sites that
differ, so the model predicts $\Jt_{\mathrm{ring}}(\text{stripe2}) =
\Jt_{\mathrm{ring}}(\text{stripe1})$ with no free parameter. All configurations are
insulating and the bilinear sector stays clean. The five values, spanning $7$~meV
[Fig.~\ref{fig:tower}(b)], are reproduced by
\begin{center}
$\Jring = 29.57$, \quad $J^{(6)}_{\mathrm{e}} = 3.49$, \quad
$J^{(6)}_{\mathrm{zz}} = 0.90$, \quad $J^{(8)} = 0.57$~meV,
\end{center}
and the prediction holds: stripe2 returns $29.0020$~meV against $29.0007$~meV, a
$1.3$~$\mu$eV deviation. The full data are collected in the Supporting Information.

The all-electron code reads the same decomposition across methods. Since FM and
N\'eel share $\Gamma_{2}^{(\mathrm{zz})} = \Gamma_{4} = +1$ and differ only in
$\Gamma_{2}^{(\mathrm{e})}$, half their difference is the domino amplitude and half their
sum the combination $\Jring + J^{(6)}_{\mathrm{zz}} + J^{(8)}$ both baths see alike:
WIEN2k gives $J^{(6)}_{\mathrm{e}} = 6.45$~meV against VASP's $3.49$, while the shared
combination agrees to $4\%$ ($29.90$ against $31.03$~meV). The loop amplitudes, small
residuals of near-cancelling energies, are the code-sensitive part; the bare ring physics
is not. The amplitudes form a converging tower [Fig.~\ref{fig:tower}(c)], each order about
$(t/U)^2$ below the last, as the counting requires: the reference dependence of
Table~\ref{tab:16state} is not a failure of the construction but its sensitivity,
resolving the bare $\Jring = 29.57$~meV from the loop corrections any local quadrilinear
probe must see.

\paragraph{Supercell design: aliasing of the pair shells.}
A practical warning follows, and it applies to energy mapping generally. In the
$2\times2\times1$ cell the third in-plane neighbor, $J_4$ at $2a = 7.5634$~\AA, is
\emph{the atom's own periodic image}: every such bond contributes $\sigma_i^2 = +1$,
the column is the constant $+16$, and $J_4$ is exactly collinear with $E_0$, so that it is
silently absorbed into the constant. The fit is then perfect (8.6~$\mu$eV) for a spurious reason.
In the $2\times3\times1$ cell $J_4$ is nominally determinable, but the $8.4562$~\AA{}
shell is \emph{exactly} linearly dependent on the lower ones, so the fitted $J_4$
absorbs it; omitting $J_4$ altogether inflates the RMS from 0.70 to 2.77~meV and leaves
residuals that track the symmetry classes rather than scattering randomly. At least
three repetitions along \emph{both} in-plane axes are needed to separate these shells.
A related sensitivity was noted by Fedorova \latin{et al.},\cite{fedorova2018} who found
that adding a single configuration to their overdetermined system shifted the interplane
couplings by up to $25\%$, and who consequently assigned a $\pm25\%$ uncertainty to their
extracted exchanges. The lesson is the same in both cases: the bilinear couplings
obtained by energy mapping are delicate objects, sensitive to the configuration set and
to the shells the supercell can resolve.

Crucially, $\Jring$ is almost untouched by all of this: it moves by $1.4\%$ between the
two cells, and by $1.4\%$ on adding $J_4$ to the $2\times3\times1$ fit, while $J_1$
moves by 1.1~meV and $J_4$ appears from nothing. The four-spin coupling is protected
because its column, being a product of four spins, is orthogonal to every bilinear
column, the same orthogonality that underlies Eq.~\eqref{eq:extract}.

\paragraph{SrFeO$_2$.}
The infinite-layer ferrous oxide provides a second test at higher spin ($S = 2$).
Symmetry analysis of the Fe sublattice gives an out-of-plane coupling $J_1$ along
$\mathbf{c}$ (3.4580~\AA), the in-plane nearest-neighbor coupling $J_2$ (3.9850~\AA), an
inter-layer diagonal $J_3$ (5.2762~\AA), and the in-plane next-nearest-neighbor $J_4$
(5.6356~\AA). In the language of the derivation above, $J_2$ is the plaquette edge and $J_4$ the
plaquette diagonal; the shell numbering, being by distance, does \emph{not} coincide
with that of La$_2$CuO$_4$. $J_1$ and $J_3$ both leave the FeO$_2$ plane, connect the
plaquette only to the bath, and cancel identically in Eq.~\eqref{eq:extract}
(Supporting Information). Results are collected in Table~\ref{tab:sfo}, alongside the
values obtained for the same four shells by Xiang, Wei and Whangbo\cite{xiang2008} from
a five-configuration energy mapping at $U_{\rm eff} = 4.6$~eV. Once their shell labels
are mapped onto ours by distance, the two determinations agree throughout, down to the
sign and magnitude of the small inter-layer diagonal coupling, $-0.23$ against our
$-0.22$~meV, and their in-plane coupling of $7.04$~meV coincides with our ring-corrected
four-state value of $7.04$~meV.

\begin{figure}[htb]
\centering
\includegraphics[width=\linewidth]{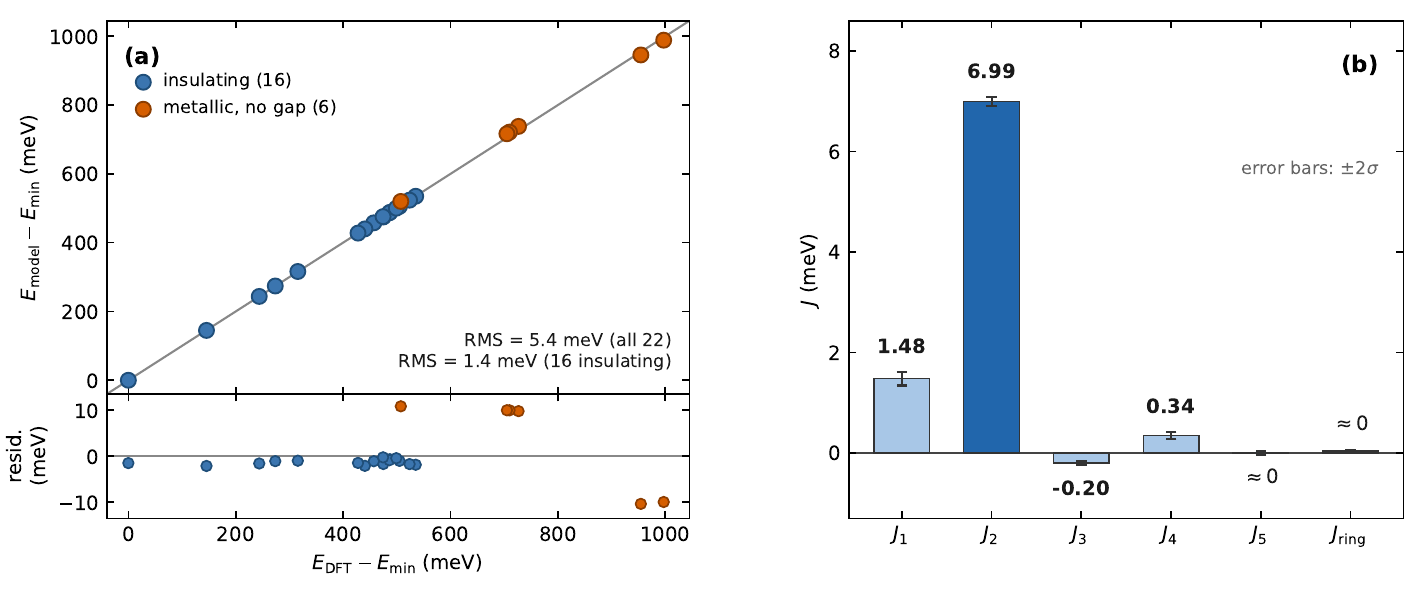}
\caption{Energy mapping for SrFeO$_2$: $2\times2\times2$ supercell,
$10\times10\times10$ $k$-mesh, 22 inequivalent collinear configurations, tetrahedron
integration. (a) Model energies against DFT energies, both referred to the lowest
configuration, with the residuals below. The six configurations whose gap closes
(Table~\ref{tab:diag}) are drawn in orange, the sixteen insulating ones in blue: the
metallic configurations carry all the large residuals, near $\pm 10$~meV where every
insulating configuration lies within $\pm 2.1$~meV (RMS $1.4$~meV for the blue points);
a refit restricted to the sixteen insulating configurations reaches $0.50$~meV (see
text). The origin of the poorer SrFeO$_2$ fit is thus visible at
a glance: gap closure, not a missing term in the Hamiltonian. (b) The extracted
couplings, with $\pm 2\sigma$ error bars. The in-plane nearest-neighbor exchange $J_2$
(the plaquette edge) dominates at $6.99$~meV; the out-of-plane $J_1$ is $1.48$~meV;
$J_3$, $J_4$ and $J_5$ are small; and the four-spin ring coupling is negligible on this
scale, $\Jring = 0.040$~meV, i.e.\ $\Jring/J_2 = 0.006$. In La$_2$CuO$_4$ the same
ratio is $0.25$. SrFeO$_2$ therefore possesses the four-site plaquette but not the ring
exchange it could support, a reminder that a plaquette is necessary for $\Jring$ but
far from sufficient.}
\label{fig:sfo}
\end{figure}

The energy mapping is shown in Fig.~\ref{fig:sfo}. The in-plane nearest-neighbor
coupling dominates, $J_2 = 6.99$~meV, with the out-of-plane $J_1 = 1.48$~meV four times
smaller, and $J_3$, $J_4$ and $J_5$ small or negligible. The four-spin ring coupling is
$\Jring = 0.040$~meV, which is to say \emph{negligible on the scale of the pair exchange}:
\begin{equation}
\Jring/J_2 = 0.006 \;\;(\text{SrFeO}_2),
\qquad\text{against}\qquad
\Jring/J_1 = 0.25 \;\;(\text{La}_2\text{CuO}_4).
\label{eq:contrast}
\end{equation}
The contrast is a factor of forty in the coupling constants, though not in the
energies. It is $J S^x$, not $J$, that enters $E_{\rm spin}$: the pair energies
$J S^2$ are nearly equal in the two materials and the ring energies $\Jring S^4$
differ by only a factor of three, as quantified below. The Fe$_4$ plaquette is
there, and the sixteen-state construction applies to it verbatim; but a plaquette is a
\emph{necessary} condition for ring exchange, not a sufficient one. Ring exchange is a
fourth-order process in the transfer integral $t$ between neighboring magnetic sites,
$\Jring \propto t^4/U^3$ with $U$ the on-site Coulomb repulsion, whereas the pair
exchange is only of second order, $J \propto t^2/U$, so the explanation that first comes
to mind is a weak hybridization. It is not the right one here. The electronegativity
difference with oxygen is nearly the same for Fe as for Cu, and the local moments show
that the covalency is comparable: the integrated Cu moment of La$_2$CuO$_4$ is
$0.7$--$0.8\,\mu_B$ against the nominal $1\,\mu_B$ and the Fe moment of SrFeO$_2$ is
$3.66\,\mu_B$ against $4$, so that in absolute terms the two ions delocalize a similar
amount of spin onto their ligands, $0.2$--$0.3\,\mu_B$ each. The couplings say the same.
The pair-exchange energy $J S^2$, which measures the hybridization summed over the
exchange channels of a bond, is nearly equal in the two materials, $28.0$ against
$32.6$~meV, and even the quadrilinear energy $\Jring S^4$ is suppressed only by a factor
of three, $0.65$ against $2.04$~meV. Hybridization alone cannot produce a factor of
forty.

The difference lies instead in the orbital structure of the magnetic center, and in the
spin it carries. In La$_2$CuO$_4$ the Cu$^{2+}$ ion ($d^9$, $S = 1/2$) holds a single
magnetically active orbital, $d_{x^2-y^2}$, $\sigma$-bonded to the bridging oxygens: the
entire covalency of the ion is concentrated in the one orbital that circulates, and the
electron travels through the same orbital at every corner of the plaquette, by four
identical $\sigma$-type steps whose amplitudes add coherently. In the high-spin
Fe$^{2+}$ ion ($d^6$, $S = 2$) of SrFeO$_2$ the configuration is
$(d_{z^2})^2(d_{xz}d_{yz})^2(d_{xy})^1(d_{x^2-y^2})^1$:\cite{xiang2008,rahman2013} the
same overall covalency is spread over four magnetic orbitals and inequivalent exchange
channels, $d_{x^2-y^2}$ through the oxygen $p_\sigma$ and $d_{xy}$ through
$p_\pi$,\cite{rahman2013} and Hund coupling locks any circulating electron to the
$S = 3/2$ core it leaves behind, whereas the Cu electron leaves behind a closed shell,
a core of spin zero, and circulates unconstrained. Pair exchange is indifferent to this dilution, since it
simply sums over the orbital channels of each bond, which is why $J S^2$ comes out
nearly the same in the two materials. The cyclic process is not indifferent: it requires
a single electron to complete four coherent hops within one orbital channel around the
loop, and essentially only the $\sigma$ channel qualifies; this is the modest factor of
three by which $\Jring S^4$ falls. The remaining factor of sixteen in $\Jring/J$ is the
classical normalization, $S^2 = 4$ against $1/4$: the coupling constants divide
spin-invariant energies by powers of $S$, so the very quadrilinear energy that yields
$\Jring = 33$~meV for a spin $1/2$ would yield only $2$~meV for a spin $2$. The two
factors, one electronic and one of pure spin bookkeeping, act in the same direction, and
together they produce the factor of forty.

The ring renormalization of the pair coupling is nevertheless still detectable, and still
in the predicted direction. The four-state calculation on a N\'eel reference returns
$J_2 = 6.72$~meV against $6.99$~meV from the energy mapping, a deficit of $0.27$~meV;
adding the predicted $2\Jring S^2 = 0.32$~meV gives $7.04$~meV, recovering the mapping
value to $0.05$~meV. Inverting the deficit as in Eq.~\eqref{eq:jring_from_contam} gives
$\Jring = 0.034$~meV, against $0.040$~meV from the mapping. The correction is
proportionally far smaller here than in the cuprate, $4\%$ of $J_2$ rather than $12\%$ of
$J_1$, but it acts with the predicted sign and magnitude at a spin value four times
larger.

\paragraph{Why the SrFeO$_2$ mapping is poorer: gap closure, not a missing term.}
One caveat must be recorded, and it proves to be the most instructive
single observation of the SrFeO$_2$ study. The fit is markedly poorer
than that of the cuprate: RMS $= 5.4$~meV over 22 configurations
[Fig.~\ref{fig:sfo}(a)], with individual residuals reaching $\pm
11$~meV, against $8.6$~$\mu$eV for La$_2$CuO$_4$. The scatter is
\emph{not} a missing biquadratic term, which is a constant on any
collinear state and therefore cannot produce residuals at all. Its
origin is written plainly in the per-configuration band gaps that
\textsc{Mag4} reports alongside every energy (Table~\ref{tab:diag}). Six
of the 22 configurations, the fully ferromagnetic one among them, are
\emph{metallic}, their gap closing entirely, whereas the remaining
sixteen keep gaps of $0.22$ to $0.83$~eV and La$_2$CuO$_4$ stays gapped
throughout, from $0.58$~eV for its ferromagnetic configuration to
$1.89$~eV for the N\'eel one. The local moments, tellingly, give no such
warning: every SrFeO$_2$ configuration retains a well-formed Fe moment,
$|\mu| \approx 3.66\,\mu_B$ per site, uniform to $0.04\,\mu_B$ across the
entire set and unchanged whether the run uses the tetrahedron method or
Gaussian smearing, so that a moment check alone passes all 22. (The
metallic states betray themselves instead in the \emph{total}
magnetization, which is non-integer, $31.71\,\mu_B$ for the
ferromagnetic configuration against the nominal $32$, as it must be
when $E_F$ crosses a band.) It is the
vanishing gap, not a collapsing moment, that marks a configuration as
having left the localized-spin regime that Eq.~\eqref{eq:ham} presumes:
an itinerant magnetic state carries energy in channels that no sum of
pairwise spin products can represent, even while each ion still holds its
spin. The correlation with the fit is then exact. The six metallic
configurations are precisely the six whose residuals exceed $\pm 9$~meV,
while all sixteen gapped configurations lie within $\pm 2.1$~meV, an RMS
of $1.4$~meV at the full-fit couplings; refitting on the sixteen gapped
configurations alone collapses the RMS further, to $0.50$~meV, a factor
of ten below the full-set value. This confirms directly that the
scatter is a physical signal and not numerical noise, and the refit
validates the couplings themselves: every parameter the gapped subset
can resolve is unchanged while its uncertainty shrinks,
$J_1 = 1.488 \pm 0.021$ against $1.482 \pm 0.138$~meV and
$J_3 = -0.196 \pm 0.010$ against $-0.199 \pm 0.044$~meV.

The subset carries a lesson of its own, however. With the six metallic
configurations removed, the fit columns of $J_2$, $J_4$ and $\Jring$
become exactly collinear (\textsc{Mag4} flags the degeneracy through
infinite variance-inflation factors), so the three can no longer be
separated; only two combinations remain determined, and both agree with
the full fit, $(J_2 - 2J_4)S^2 = 25.2$ against $25.2$~meV and
$J_4 S^2 - \Jring S^4 = 0.74$ against $0.73$~meV. The high-magnetization
configurations, which are exactly the ones that go metallic, are also
exactly the ones that lift this degeneracy. One therefore cannot simply
discard the suspect configurations after the fact and keep the full
coupling model, a configuration-set analogue of the supercell aliasing
discussed below; the SrFeO$_2$ \emph{pair} couplings thus carry a larger
uncertainty than their formal error bars suggest, while $\Jring$, whose
column over the \emph{full} configuration set is orthogonal to every
bilinear one, is protected in the full fit but not resolvable from the
gapped subset alone.

The four-state method fails on the ferromagnetic reference in the same way, and
the diagnostics of Table~\ref{tab:diag} say why, turning what would otherwise be an
unexplained outlier into the clearest illustration of the reference-choice argument
developed above. In the original run ($3\times3\times3$ cell,
$3\times3\times3$ $k$-mesh) the corrected $J_2 = 5.83$~meV lies $1.2$~meV below the
mapping value and $J_3 = +0.61$~meV carries the wrong sign; repeating it at a denser
$5\times5\times5$ $k$-mesh changes nothing, $J_2 = 5.78$~meV and $J_3 = +0.70$~meV,
because the problem is not one of numerical convergence. All five configurations of
the ferromagnetic-reference set are metallic, while their local Fe moments remain
perfectly healthy at $3.59$--$3.69\,\mu_B$: the reference has crossed the metallic
edge, and every coupling extracted on it is compromised. The symptoms are exactly
the predicted ones. The out-of-plane $J_1$, which belongs to no plaquette and must
therefore be reference-independent, shifts from $1.55$~meV (N\'eel) to $2.36$~meV
(FM); and the two-bath formula of Eq.~\eqref{eq:twobath}, fed with this reference,
returns $\Jring = (6.10 - 6.72)/4S^2 = -0.04$~meV, wrong even in sign against the
$+0.04$~meV of the mapping. The two-bath route is therefore simply not available in
SrFeO$_2$: it requires a trustworthy ferromagnetic reference, and the material
declines to provide one. The N\'eel-reference values, by contrast, reproduce the
mapping $J_2$ to $0.11$ and $0.05$~meV, and do so consistently in two supercells
of different shape and size, both built on the $\sqrt2$ in-plane cell, 24 Fe
($2\times2\times3$) and 54 Fe ($3\times3\times3$): every configuration of both sets
is gapped, at $0.44$ to $0.72$~eV in the larger cell, small but nonzero, which is
again the distinction that matters, and the two cells agree, $J_1 = 1.557$ against
$1.554$, $J_2 = 6.774$ against $6.717$, and $J_3 = -0.222$ against $-0.222$~meV.
This is the practical content of the recommendation made with the
diagnostics: in a gap-sensitive material, anchor the local methods on the gapped,
ground-state-like reference.

\begin{table}[htb]
\caption{Band gap and local-moment diagnostics printed by \textsc{Mag4} for every
configuration. (a) Energy-mapping sets: both materials keep well-formed moments
throughout, so the gap is the discriminating quantity. La$_2$CuO$_4$ is gapped on all 13 configurations (mapping exact); six of the 22 SrFeO$_2$ configurations are metallic and carry all the large residuals, and removing them drops the RMS tenfold. (b) Four-state sets, with the raw (uncorrected) plaquette-edge coupling: $J_1$ for La$_2$CuO$_4$ ($2\times2\times1$, $9\times9\times5$), $J_2$ for SrFeO$_2$ (FM: $3\times3\times3$, 27 Fe, $5\times5\times5$; N\'eel: $\sqrt2$-cell $3\times3\times3$, 54 Fe, $3\times3\times5$). $^{a}$Moments are stable throughout, $|\mu|\approx 3.7\,\mu_B$ per Fe and $0.7$--$0.8\,\mu_B$ per Cu.}
\label{tab:diag}
\centering
\small
%
\begin{tabular}{l c c c c c}
\toprule
\multicolumn{6}{l}{\emph{(a) Energy-mapping configuration sets}}\\
Material & $N_{\rm cfg}$ & \# metallic & gap range$^{\ast}$ (eV)
  & moments & RMS (all\,$\to$\,gapped) \\
\midrule
La$_2$CuO$_4$ ($2\times2\times1$) & 13 & 0 & 0.578--1.885
  & all pass$^{a}$ & 0.0086~meV \\
SrFeO$_2$ ($2\times2\times2$) & 22 & 6 & 0.220--0.833
  & all pass$^{a}$ & 5.4\,$\to$\,0.50~meV \\
\bottomrule
\end{tabular}

\vspace{2pt}
{\footnotesize $^{\ast}$range over the gapped configurations; the
metallic ones have zero gap.}

\vspace{8pt}
%
\begin{tabular}{l l c c l r}
\toprule
\multicolumn{6}{l}{\emph{(b) Four-state configuration sets (plaquette-edge coupling, raw)}}\\
Material & Reference & $N_{\rm cfg}$ & gap range (eV) & moments & $J_{\rm edge}$ (meV) \\
\midrule
La$_2$CuO$_4$ & FM & 4 & 0.578--0.596 & all pass$^{a}$ & 146.83 \\
La$_2$CuO$_4$ & N\'eel        & 4 & 0.596--1.885 & all pass$^{a}$ & 114.19 \\
SrFeO$_2$     & FM & 5 & all metallic & all pass$^{a}$ & 6.10 \\
SrFeO$_2$     & N\'eel        & 5 & 0.435--0.724 & all pass$^{a}$ & 6.72 \\
\bottomrule
\end{tabular}
\end{table}

\begin{table}[htb]
\caption{Exchange couplings of SrFeO$_2$ (meV), $S = 2$. Four-state values are quoted
before and after the ring correction of Eq.~\eqref{eq:contam}. The last column lists the
LDA$+U$ ($U_{\rm eff} = 4.6$~eV) energy-mapping values of Xiang, Wei and
Whangbo,\cite{xiang2008} obtained in the same Hamiltonian convention as
Eq.~\eqref{eq:ham}; their shells, labeled in a different order, are mapped onto ours by
distance (their $J_1$, $J_2$, $J_3$, $J_4$ are our $J_2$, $J_1$, $J_4$, $J_3$), and they
report that $U$ values between 3 and 6~eV lead to qualitatively the same results.}
\label{tab:sfo}
\centering
\small
\begin{tabular}{l c r r r}
\toprule
Coupling & $d$ (\AA) & mapping & 4-state (N\'eel) & Ref.~\citenum{xiang2008} \\
\midrule
$J_1$ (out-of-plane $\parallel\mathbf{c}$) & 3.4580 & 1.482 & 1.554 & 2.18 \\
$J_2$ (in-plane NN = plaq. edge)           & 3.9850 & 6.989 & 6.717 & 7.04 \\
$J_2$ + ring correction                    &        & ---   & \textbf{7.040} & \\
$J_3$ (inter-layer diagonal)               & 5.2762 & $-0.199$ & $-0.222$ & $-0.23$ \\
$J_4$ (in-plane NNN = plaq. diag.)          & 5.6356 & 0.345 & --- & 0.43 \\
$J_5$                                       & ---    & $-0.004$ & --- & \\
$\Jring$                                    & ---    & 0.0404 & --- & \\
$\Jring$ (sixteen-state)$^{d}$              & ---    & \multicolumn{2}{c}{0.0526} & \\
\midrule
$\Jring/J_2$                                &        & 0.0058 & & \\
fit RMS (meV)                               &        & 5.4    & & \\
\bottomrule
\end{tabular}

\vspace{2pt}
{\footnotesize $^{d}$ N\'eel reference, isolated plaquette in a $\sqrt2$-cell $2\times2\times2$ supercell (16 Fe); a $2\times2\times1$ cell gives
$0.074$~meV, and two of the eight configurations are metallic (see Supporting
Information).}
\end{table}

\paragraph{Code comparison: VASP versus WIEN2k.}
All results above use the plane-wave PAW method. To check that they are not an artefact
of the pseudopotential construction, the same couplings were computed with the
all-electron full-potential linearized augmented-plane-wave code
WIEN2k,\cite{blaha2020} at the same $U_{\rm eff}$ (Table~\ref{tab:codes}). There, the muffin-tin radii
were chosen to coincide with the PAW projection radii used for the DFT+U
correction in VASP, so that the on-site term acts on comparable atomic
spheres in both methods. The $U$ correction nonetheless acts on different radial wavefunctions in the two codes, so
exact agreement is not expected.

For La$_2$CuO$_4$ the two codes agree to within a few per cent on every quantity that
can be compared. The nearest-neighbor coupling is $130.5$~meV in VASP (energy mapping)
against $132.3$--$137.0$~meV in WIEN2k, and the sixteen-state $\Jring$ on a
ferromagnetic reference is $34.5$~meV in VASP against $36.4$~meV in WIEN2k, a $5\%$
difference. The WIEN2k reproduction of the sixteen-state reference dependence,
and its reading in terms of the loop amplitudes, were presented with
Table~\ref{tab:16state} and Eq.~\eqref{eq:tower} above; two further checks belong
here. The ferromagnetic sixteen-state value moves by only $0.02$~meV between the
coarse and the dense mesh. And a least-squares fit of the same WIEN2k energies with
$\Jring$ included returns $J_1 = 134.0$--$134.8$~meV, bare as the derivation
requires, within $3\%$ of the VASP mapping value.

Note that the WIEN2k
$1\times1\times2$ cell is one in which $J_1$ carries \emph{no} ring term at all, so its
FM and N\'eel values are uncorrected and ought to coincide; that they differ by
$4.7$~meV reflects their different $k$-meshes ($12\times6\times6$ against
$6\times3\times3$) rather than any ring physics, and a matched-mesh pair is being
computed. For SrFeO$_2$ the $J_2$ values from WIEN2k are somewhat smaller than
the VASP ones, but internally consistent between two different supercells, one that
carries no ring correction and one ($2\times2\times3$) that does; the residual
offset against VASP could reflect the different projection of the on-site $U$ in
the two methods.

\begin{table}[htb]
\caption{Code comparison at matched $U_{\rm eff}$. Four-state values are quoted after
the ring correction of Eq.~\eqref{eq:contam}; in the WIEN2k $1\times1\times2$ cells
$J_1$ (resp.\ $J_2$) carries no ring term, so those entries are uncorrected.}
\label{tab:codes}
\centering
\small
\begin{tabular}{l l l c r}
\toprule
 & Code & Supercell / $k$-mesh & Ref. & Value (meV) \\
\midrule
\multicolumn{5}{l}{\emph{T-La$_2$CuO$_4$, $J_1$}}\\
& VASP   & $2\times2\times1$ / $9\!\times\!9\!\times\!5$   & ---    & 130.51 (mapping) \\
& VASP   & $2\times2\times1$ / $9\!\times\!9\!\times\!5$   & N\'eel & 130.54 \\
& WIEN2k & $1\times1\times2$ / $12\!\times\!6\!\times\!6$  & FM     & 132.30 \\
& WIEN2k & $1\times1\times2$ / $6\!\times\!3\!\times\!3$   & N\'eel & 137.02 \\[2pt]
\multicolumn{5}{l}{\emph{T-La$_2$CuO$_4$, $\Jring$ (sixteen-state)}}\\
& VASP   & $\sqrt2$-cell $2\times2\times1$ / $6\!\times\!6\!\times\!5$ & N\'eel & 27.54 \\
& VASP   & $\sqrt2$-cell $2\times2\times1$ / $6\!\times\!6\!\times\!5$ & FM & 34.52 \\
& WIEN2k & same cell$^{b}$ / $6\!\times\!5\!\times\!6$ & N\'eel & 23.45 \\
& WIEN2k & same cell$^{b}$ / $3\!\times\!3\!\times\!3$   & FM     & 36.33 \\
& WIEN2k & same cell$^{b}$ / $6\!\times\!5\!\times\!6$ & FM & 36.35 \\[2pt]
\multicolumn{5}{l}{\emph{SrFeO$_2$, $J_2$}}\\
& VASP   & $2\times2\times2$ / $10\!\times\!10\!\times\!10$& ---    & 6.99 (mapping) \\
& VASP & $\sqrt2$-cell $2\times2\times3$ / $5\!\times\!5\!\times\!5$ & N\'eel & 7.10 \\
& VASP   & $\sqrt2$-cell $3\times3\times3$ / $3\!\times\!3\!\times\!5$ & N\'eel & 7.04 \\
& VASP   & $3\times3\times3$ / $3\!\times\!3\!\times\!3$   & FM     & 5.83$^{c}$ \\
& VASP   & $3\times3\times3$ / $5\!\times\!5\!\times\!5$   & FM     & 5.78$^{c}$ \\
& VASP & $\sqrt2$-cell $2\times2\times3$ / $5\!\times\!5\!\times\!5$ & FM & 5.96$^{c}$ \\
& WIEN2k & $\sqrt2$-cell $2\times2\times3$ / $3\!\times\!3\!\times\!3$   & N\'eel & 6.49 \\
& WIEN2k & $\sqrt2$-cell $1\times1\times2$ / $12\!\times\!12\!\times\!10$& N\'eel & 6.59 \\
\bottomrule
\end{tabular}

\vspace{2pt}
{\footnotesize $^{b}$ the same sixteen-Cu isolated-plaquette supercell, indexed
$2\times1\times2$ in the WIEN2k axis convention; $k$-meshes as indexed in that
convention.} \quad $^{c}$ unreliable: all configurations of the ferromagnetic
reference are metallic (Table~\ref{tab:diag}), and the same runs return
$J_3 = +0.61$, $+0.70$ and $+0.85$~meV against $-0.20$~meV from every other
calculation; see text.
\end{table}

\paragraph{Convention.}
A caveat is worth stating, since it is a recurrent source of confusion. The
spin-product $\Jring$ of Eq.~\eqref{eq:ham} (and of Moreira \latin{et
al.}\cite{moreira2006}) differs from the Dirac cyclic-permutation amplitude $J_4$ of the multiple-spin-exchange and neutron-scattering
literature:\cite{roger1983,toader2005} expanding the permutation operator
generates, besides the four-spin term, two-spin contributions that renormalize the
effective nearest- and next-nearest exchange. Values in the two conventions must be
converted before comparison. Equation~\eqref{eq:contam} is the corresponding statement
at the level of the energy mapping, and Eq.~\eqref{eq:jring_from_contam} shows it holds
quantitatively.

\begin{figure}[htb]
\centering
\includegraphics[width=0.65\linewidth]{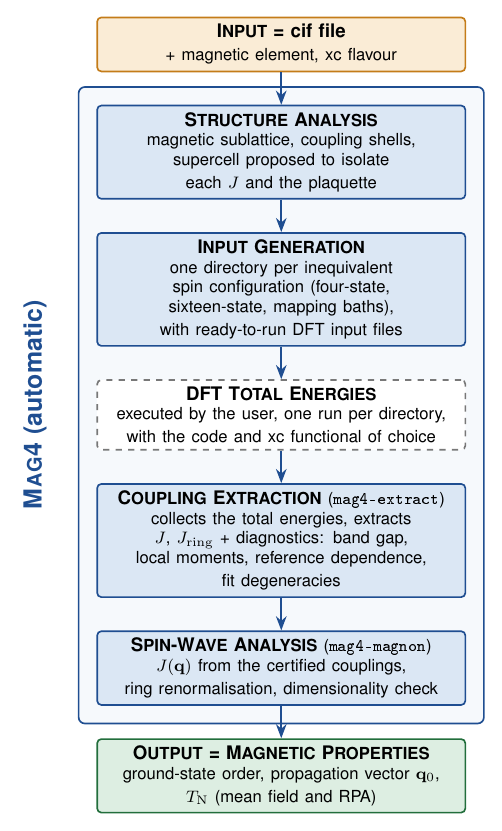}
\caption{The \textsc{Mag4} workflow, from a cif file to the predicted magnetic
properties. The only required input is the crystallographic structure, together with
the choice of magnetic element and exchange-correlation flavour; the boxed stages run
automatically. The workflow proposes the supercell that isolates each coupling,
generates every spin configuration needed by the four-state, sixteen-state, and
energy-mapping extractions, certifies each coupling through its band-gap and
local-moment diagnostics, and performs the spin-wave analysis (ring renormalisation
of the effective bilinear couplings, dimensionality check) that delivers the
predicted ground-state order, its propagation vector, and the
critical temperature in mean field and in the RPA. The dimensionality check warns
when the extracted couplings cannot, by themselves, order the material, as in
T-La$_2$CuO$_4$.}
\label{fig:workflow}
\end{figure}

\paragraph{From a cif file to the magnetic order and its critical
temperature: the Mag4 workflow.}
The extraction described so far is one stage of a chain that \textsc{Mag4} automates
from end to end (Fig.~\ref{fig:workflow}). The only required input is a
crystallographic cif file: the package identifies the magnetic sublattice, proposes
the supercell that isolates each coupling, and generates one ready-to-run directory of
DFT input files per inequivalent spin configuration, in the exchange-correlation
flavour the user selects. Once the user has executed these calculations,
\textsc{Mag4} collects the total energies and returns the couplings certified by
their band-gap and local-moment diagnostics.

From the couplings, \textsc{Mag4} then predicts what the chemist actually wants
to know: which magnetic order the material adopts, and at which temperature. The
ground state and its propagation vector follow from the Luttinger-Tisza
construction,\cite{luttinger1946} the critical temperature from mean field and from
the random phase approximation (RPA, Tyablikov decoupling\cite{bogolyubov1959}), with
the ring exchange folded automatically into the effective bilinear
couplings.\cite{coldea2001} For SrFeO$_2$ the prediction is the G-type order observed
by neutron diffraction, $\mathbf{q}_0 = (1/2, 1/2, 1/2)$, with
$T_{\mathrm{N}}^{\mathrm{RPA}} = 447$~K against the experimental
$473$~K:\cite{tsujimoto2007} within six percent. For T-La$_2$CuO$_4$ the predicted
in-plane vector is the N\'eel order of the CuO$_2$ plane, but the workflow detects
that the interlayer coupling vanishes (it is exactly frustrated in the body-centered
tetragonal cell) and, invoking the Mermin-Wagner theorem,\cite{mermin1966} reports its
$461$~K not as a N\'eel temperature but as the in-plane correlation scale. The real
material orders at $325$~K through what the ideal cell omits, the orthorhombic
distortion that releases an effective interlayer coupling and the small
anisotropies,\cite{kastner1998,coldea2001} terms that lie beyond the isotropic
collinear model considered here.

Two remarks keep these numbers in perspective. First, the couplings come from
DFT$+U$, which is not exact: numerical accuracy matters, and SrFeO$_2$ shows how far
it can go, but for complex materials the trends matter more, since they are what
reveal the driving forces and suggest routes to improve a property; a tool that
returns the right order, the right scale, and the right ranking is what a chemist
needs to decide which candidate deserves synthesis. Second, a screening tool must
know the limits of its own model: the same run that certifies each coupling also
states, as for T-La$_2$CuO$_4$, whether those couplings can by themselves order the
material.

\section{Conclusions}

We have introduced a sixteen-state energy-mapping scheme that isolates the four-spin
ring coupling $\Jring$ from collinear GGA$+U$ total energies. The alternating sum over the
sixteen collinear arrangements of a nearest-neighbor plaquette cancels the
spin-independent constant, the molecular fields of the bath, and every pair coupling
exactly, for any choice of reference. Symmetry reduces the sixteen configurations to a few inequivalent energies in the magnetic (grey) group: six on the ideal single-layer lattice
and eight in the body-centered stacking of T-La$_2$CuO$_4$; the further degeneracies the spin model predicts, but symmetry does not enforce, are confirmed by the DFT energies to $10$~$\mu$eV.

The derivation also shows that the conventional four-state pair coupling is
ring-renormalized, with a sign set by the reference, and T-La$_2$CuO$_4$ bears this out:
three routes resting on different energy differences agree on $\Jring = 32.6$--$32.7$~meV
to $0.2\%$, giving $\Jring/J_1 = 0.25$ in line with periodic-DFT and experimental
estimates, while a four-state $J_1$ quoted without naming its reference is wrong by
$12\%$. The four-spin coupling is moreover remarkably insensitive to truncation of the
bilinear model, where $J_1$ is not, because its four-spin column is orthogonal to every
pair column, the same orthogonality that underlies the extraction.

The direct sixteen-state extraction adds a finding of its own. On an isolated plaquette,
with every diagnostic clean, it still depends on the reference: $27.5$~meV on a N\'eel
bath against $34.5$~meV on a ferromagnetic one. A pure pair-plus-ring model forbids this,
and the bilinear couplings stay reference-independent, so the difference is a genuine
fourth-order signature of interactions beyond that model. Turned around, it becomes a measurement: each bath weights each loop family by a
known correlator, so four inequivalent baths separate the bare $\Jring = 29.57$~meV from a
series of successively smaller six- and eight-spin loop corrections, and a fifth bath
confirms the parameter-free prediction to $1.3$~$\mu$eV.

The two test materials answer different questions. La$_2$CuO$_4$ shows that the ring term
is large and that ignoring it corrupts $J$; SrFeO$_2$ shows the converse, with $\Jring/J_2 = 0.006$. The contrast in the
couplings overstates the contrast in the physics, since it is $J S^x$ that enters the
energies: the pair energies $J S^2$ are nearly equal ($28.0$ against $32.6$~meV) and the
ring energies $\Jring S^4$ differ only by a factor of three ($0.65$ against $2.04$~meV),
the rest being the $1/S^x$ normalization.
Read together, the two materials leave a structure-property lesson: they present
the same magnetic plaquette, yet their relative ring exchange differs by a factor of
forty, so the plaquette motif is a necessary condition and nothing more. The chemical
factors that separate the two cases (the covalency of the metal-ligand bond, the
on-site repulsion, the orbital filling) are the levers on which a chemist can act, and
they turn the search for strong multi-spin magnetism into a concrete screening problem.

Because it requires only collinear single-point energies and a small, symmetry-reduced set
of configurations, the method applies readily to the growing family of two-dimensional and
correlated magnets in which higher-order exchange is suspected. The whole workflow is
implemented in the openly available \textsc{Mag4} package,\cite{mag4} so that obtaining
$\Jring$ costs no more effort than obtaining $J$.
A candidate structure, existing or hypothetical, can therefore be screened from
its cif file alone: the supercell, the DFT inputs in the chosen
exchange-correlation flavour, the couplings $J$ and $\Jring$ with their reliability
diagnostics, the predicted magnetic order with its propagation vector, and an estimate
of the critical temperature all follow automatically, and the workflow itself states
when the extracted couplings cannot, by themselves, order the material.
In this, the tool simply follows the path that Mike Whangbo has traced and continues
to trace: sophisticated physics distilled into practical instruments and placed in
the hands of chemists, to explain, to rationalize, and to predict the magnetic
materials still to be discovered.

\begin{acknowledgement}
The authors acknowledge Grand équipement national de calcul intensif (GENCI) for granting access to the High-performance computing (HPC) resources of TGCC (Très grand centre de calcul du CEA), CINES (Centre informatique national de l'enseignement supérieur) and IDRIS (Institut du développement et des ressources en informatique scientifique) networks under the allocation 2026-A0190907682.
\end{acknowledgement}

\begin{suppinfo}
The complete derivation, written out step by step and with nothing contracted: the geometry
of the plaquette and of every plaquette touching it, the collinear reduction of the ring
operator, the exact decomposition of the energy, all sixteen energies in full, the
cancellation identities, the extraction formula, the role of time reversal in the symmetry
reduction, and the proof that the four-state method returns the effective coupling rather
than the bare one. This is followed by the derivation worked through for SrFeO$_2$, with
the couplings $J_1$ (out-of-plane), $J_2$ (in-plane nearest-neighbor), $J_3$ (inter-layer
diagonal) and $J_4$ (in-plane next-nearest-neighbor) classified explicitly, and by the
supercell and convergence checks, and by the full data of the five-bath
decomposition: the bath sign patterns, correlators, gap ranges, measured and modeled
$\Jt_{\mathrm{ring}}$, and the total energies of all forty-four configurations.
\end{suppinfo}

\bibliography{refs}

\end{document}